 \documentclass[reprint, onecolumn, notitlepage, times, longbibliography, aps, footinbib, showpacs, superscriptaddress]{revtex4-1}
\usepackage[letterpaper, margin=1in]{geometry}

\usepackage[utf8]{inputenc}
\usepackage{graphicx}
\usepackage{float}
\usepackage{amsmath, braket}
\usepackage{amssymb}
\usepackage{color}
\usepackage{hyperref}
\usepackage{wrapfig}
\usepackage{subcaption}

\graphicspath{ {images/} {numerics/} }

\usepackage[dvipsnames]{xcolor}
\definecolor{Zcolour}{rgb}{0.992, 0.588, 0.22}

\newcommand{\tfd}{\textrm{TFD}}
\newcommand{\anc}{\textrm{anc}}
\newcommand{\lads}{L_{\text{AdS}}}

\begin{document}
\title{Entanglement of purification: from spin chains to holography}

\author{Phuc Nguyen}
\affiliation{Maryland Center for Fundamental Physics and Department of Physics, University of Maryland, College Park, MD 20742, USA}
\author{Trithep Devakul}
\affiliation{Department of Physics, Princeton University, Princeton, NJ 08540, USA}
\author{Matthew G. Halbasch}
\affiliation{Department of Physics, Princeton University, Princeton, NJ 08540, USA}
\author{Michael P. Zaletel}
\affiliation{Department of Physics, Princeton University, Princeton, NJ 08540, USA}
\author{Brian Swingle}
\affiliation{Condensed Matter Theory Center, Maryland Center for Fundamental Physics, Joint Center for Quantum Information and Computer Science, and Department of Physics, University of Maryland, College Park, MD 20742, USA}

\begin{abstract}

Purification is a powerful technique in quantum physics whereby a mixed quantum state is extended to a pure state on a larger system. This process is not unique, and in systems composed of many degrees of freedom, one natural purification is the one with minimal entanglement. Here we study the entropy of the minimally entangled purification, called the entanglement of purification, in three model systems: an Ising spin chain, conformal field theories holographically dual to Einstein gravity, and random stabilizer tensor networks. We conjecture values for the entanglement of purification in all these models, and we support our conjectures with a variety of numerical and analytical results. We find that such minimally entangled purifications have a number of applications, from enhancing entanglement-based tensor network methods for describing mixed states to elucidating novel aspects of the emergence of geometry from entanglement in the AdS/CFT correspondence.

\end{abstract}

\maketitle

\tableofcontents

\section{Introduction}

In quantum physics, it is always possible to interpret the entropy of a physical system as arising from entanglement with an auxiliary system. Given a physical system in a mixed quantum state, one can introduce a fictitious auxiliary system such that the combined system is in a pure state. The state of the combined system is called a purification of the original mixed state and the entanglement between the purifier and the original system, as encoded in the entanglement entropy, recovers the von Neumann entropy of the original state.

For example, when studying the thermal physics of quantum systems, it is often useful to work with a state called the thermofield double which purifies the thermal Gibbs state. In the context of numerical simulations of strongly interacting quantum spin chains using tensor network methods, the thermofield double construction is useful because it maps thermal entropy to entanglement entropy and opens up new algorithmic tools \cite{feiguin}. In the context of the AdS/CFT correspondence, the thermofield double construction is also useful and takes on an interesting physical meaning. The AdS/CFT correspondence maps thermal states to black holes \cite{Maldacena:1997re,Witten:1998qj,GUBSER1998105}, and the thermofield double state is mapped is to a wormhole geometry that connects the original black hole with a second black hole (the auxiliary system) \cite{1976PhLA...57..107I,Maldacena:2001kr}.

However, the thermofield double is only one purification of the thermal state; there all an infinite number of other purifications which are all related by the action of a unitary transformation on the auxiliary system. In the context of tensor network methods, where entanglement is a precious resource, it would be especially useful to work with a purification which had the minimal possible entanglement. It is also interesting to ask if the minimal purification has any geometric meaning within the AdS/CFT correspondence. Indeed, we expect there to be a connection between these two directions given the relationship between tensor networks and the AdS/CFT correspondence \cite{Swingle:2009bg}.

Remarkably, the notion of a purification with the minimal possible entanglement has also been considered in quantum information science as one measure of the total correlations present in a bipartite mixed state. This quantity is called the entanglement of purification \cite{2002JMP....43.4286T}, and here we study it in the context of three different classes of quantum many-body systems. We consider first a class of strongly coupled conformal field theories which are holographically dual to Einstein gravity. Next we study a spin chain whose low energy physics is described by an Ising conformal field theory. We also report a result in a random stabilizer state tensor network model \cite{Hayden:2016cfa,Nezami:2016zni}. Through a combination of analytical arguments and numerical calculations, we conjecture values for the entanglement of purification in all these systems, and, in the case of random stabilizer states, give a rigorous argument.

Our primary motivations are two fold. First, from the perspective of tensor network methods, specifically matrix product states \cite{fannes}, we want to investigate the minimal entanglement amongst purifications of a given thermal state. As indicated above, the minimal entanglement purification could be a useful technical tool in numerical simulations. Indeed, in our calculations we find that the entanglement of the thermofield double state can be reduced by as much as a factor of two, leading to a reduced bond dimension equal to the square root of the thermofield double bond dimension, a substantial reduction given a computational cost scaling like the third power of the bond dimension. Second, from the perspective of holographic models, we want to understand other geometric aspects of the bulk geometry in terms of quantum information. The Ryu-Takayanagi (RT) formula \cite{Ryu:2006bv} relating entanglement entropy to minimal surfaces is the best example of this correspondence, but it is particularly interesting to search for quantum information measures that go beyond the minimal curve paradigm and capture other aspects of the geometry.

\subsection{Technical introduction}

\begin{wrapfigure}{R}{6cm}
    \centering
    \includegraphics[width=4.5cm]{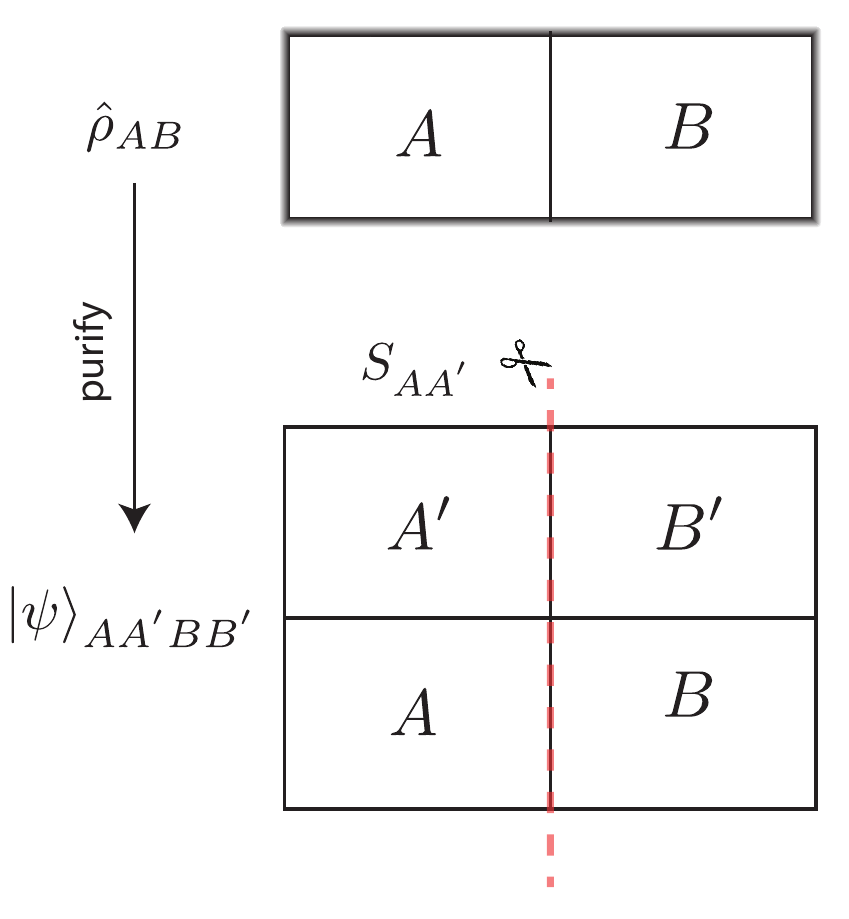}
    \caption{}
    \label{fig:PlotDefEp}
\end{wrapfigure}

The entanglement of purification (EP) is defined as follows \cite{2002JMP....43.4286T}: let $\rho_{AB}$ be a density matrix on a bipartite system $\mathcal{H}_{A} \otimes \mathcal{H}_{B}$. Let $|\psi\rangle \in \mathcal{H}_{AA'} \otimes \mathcal{H}_{BB'}$ be a purification of $\rho_{AB}$, e.g., $\mbox{Tr}_{A' B'} \ket{\psi} \bra{\psi} = \rho_{A B}$, as illustrated schematically in Figure~\ref{fig:PlotDefEp}. The EP of $\rho$ is given by:
\begin{equation}
    E_{p}{(\rho)} = \min_{\psi,A'} S_{AA'}
\end{equation}
Here we minimize over all $\psi$ and over all ways of partitioning the purification into $A'B'$, and $S_{AA'}$ is the von Neumann entropy of the reduced density matrix obtained by tracing out the $BB'$ part of $|\psi \rangle \langle \psi|$.\\

To gain some familiarity with this definition, let us consider a few simple cases. If $\rho_{AB}$ is pure,
\begin{equation}
\rho_{AB} = |\phi \rangle \langle \phi|_{AB},
\end{equation}
then no purification is needed and $E_p= S(A)=S(B)$. If $\rho_{AB}$ is uncorrelated,
\begin{equation}
\rho_{AB} = \rho_A \otimes \rho_B,
\end{equation}
then there exists a purification of the form $|\psi_1\rangle_{AA'}\otimes |\psi_2\rangle_{BB'}$ in which case $E_p=0$. If $\rho_{AB}$ is classically correlated,
\begin{equation}
\rho_{AB} = \sum_i p_i \ket{i}\bra{i}_A \otimes \ket{i}\bra{i}_B,
\end{equation}
then it can be shown that $E_p = -\sum_i p_i \log p_i$, the Shannon entropy of the $\{p_i\}$ distribution (see Appendix \ref{App:Proofs} for the proof).

More generally, $E_p$ obeys a few key properties. For readers unfamiliar with these properties, we have for completeness included proofs of these properties drawn from the literature \cite{2002JMP....43.4286T,2015PhRvA..91d2323B} in Appendix \ref{App:Proofs}.
\begin{itemize}
    \item The $E_{p}$ is bounded above by the entanglement entropy:
    \begin{equation}\label{EPUpperBound}
        E_{p}{(A:B)} \leq \mathrm{min}{(S{(A)},S{(B)})}
    \end{equation}

    \item The $E_{p}$ is monotonic, i.e. it never increases upon discarding a subsystem:
    \begin{equation}\label{Monotonicity}
        E_{p}{(A:BC)} \geq E_{p}{(A:B)}
    \end{equation}
    \item The $E_{p}$ is bounded below by half the mutual information:
    \begin{equation}\label{EPLowerBound}
        E_{p}{(A:B)} \geq \frac{I{(A:B)}}{2}
    \end{equation}
    \item For a tripartite system, we have the bound:
    \begin{equation}\label{EPTripartiteBound}
        E_{p}{(A:BC)} \geq \frac{I{(A:B)}}{2} + \frac{I{(A:C)}}{2}
    \end{equation}
    \item In a bipartite state that saturates the Araki-Lieb inequality, $S(AB) = |S(A) - S(B)|$, we have $E_{p}{(A:B)} = \mathrm{min}{(S{(A)},S{(B)})}$.
    \item For a tripartite pure state, the $E_{p}$ is polygamous:
    \begin{equation}\label{EPPolygamy}
        E_{p}{(A:B)} + E_{p}{(A:C)} \geq E_{p}{(A:BC)}
    \end{equation}
\end{itemize}

We now proceed to study the entanglement of purification in the aforementioned three classes of physical systems. In the holographic models we proceed by proposing a new dictionary entry relating entanglement of purification to minimal cross-section of the ``entanglement wedge" \cite{Headrick:2014cta,Czech:2012bh,Jafferis2016} bounded by the physical boundary and the RT surface \cite{Ryu:2006bv}. More precisely, we argue that amongst the subset of purifications which have a geometrical gravity dual, the entanglement wedge cross section is the entanglement of purification. We do not show that it suffices to restrict to geometric purifications, but we give some plausibility arguments and show that our proposal obeys all the above properties of $E_p$. Throughout we denote our holographic proposal for $E_p$ by $E_{ph}$.

In the spin chain model we proceed numerically to approximately find the minimal entanglement purification. We start from the thermofield double state and succeed in removing entanglement, but we do not rigorously show that we have found the optimal purification. However, we do find that the numerical results are in remarkable accord with the holographic proposal, perhaps more even than one might expect given that one conformal field theory has central charge less than one (spin chain) while the other has very large central charge and a sparse low lying operator spectrum (AdS/CFT). Throughout we denote the output of our spin chain numerics by $\tilde{E}_p$.

Finally, we also study a tensor network model composed of random stabilizer states. In this tensor network class, all entanglement consists of either Bell pairs or ``cat states"/GHZ states. Using recent results on the GHZ content of random stabilizer tensor network states \cite{Nezami:2016zni}, we show that in this case the entanglement of purification is approximately $\frac{1}{2} I(A:B)$ on average, i.e. near the lower bound. This is so despite the fact that entanglement entropy in such states is computed using a discrete version of the RT formula.

\textit{Note:} After our holographic results were obtained and while preparing the manuscript, a very similar holographic proposal for the entanglement of purification appeared \cite{2017arXiv170809393T}.

\section{Holographic proposal}\label{Sec:Holography}

In this section we introduce and motivate our holographic prescription for entanglement of purification, denoted $E_{ph}$. We discuss the core ideas justifying our proposal and give some sample calculations in the ground state and in thermal equilibrium at non-zero temperature. Later, in Sec.~\ref{Sec:HolographyGeneral}, we discuss generalizations of our proposal to time-dependent situations and show that $E_{ph}$ obeys all the properties listed in the technical introduction. As we discuss in detail below, our proposal for the holographic dual of $E_p$ is strongly motivated by tensor network models of the AdS/CFT correspondence.

\subsection{Proposal: time-independent geometry}

\begin{figure}[H]
    \centering
    \includegraphics[width=4.5cm]{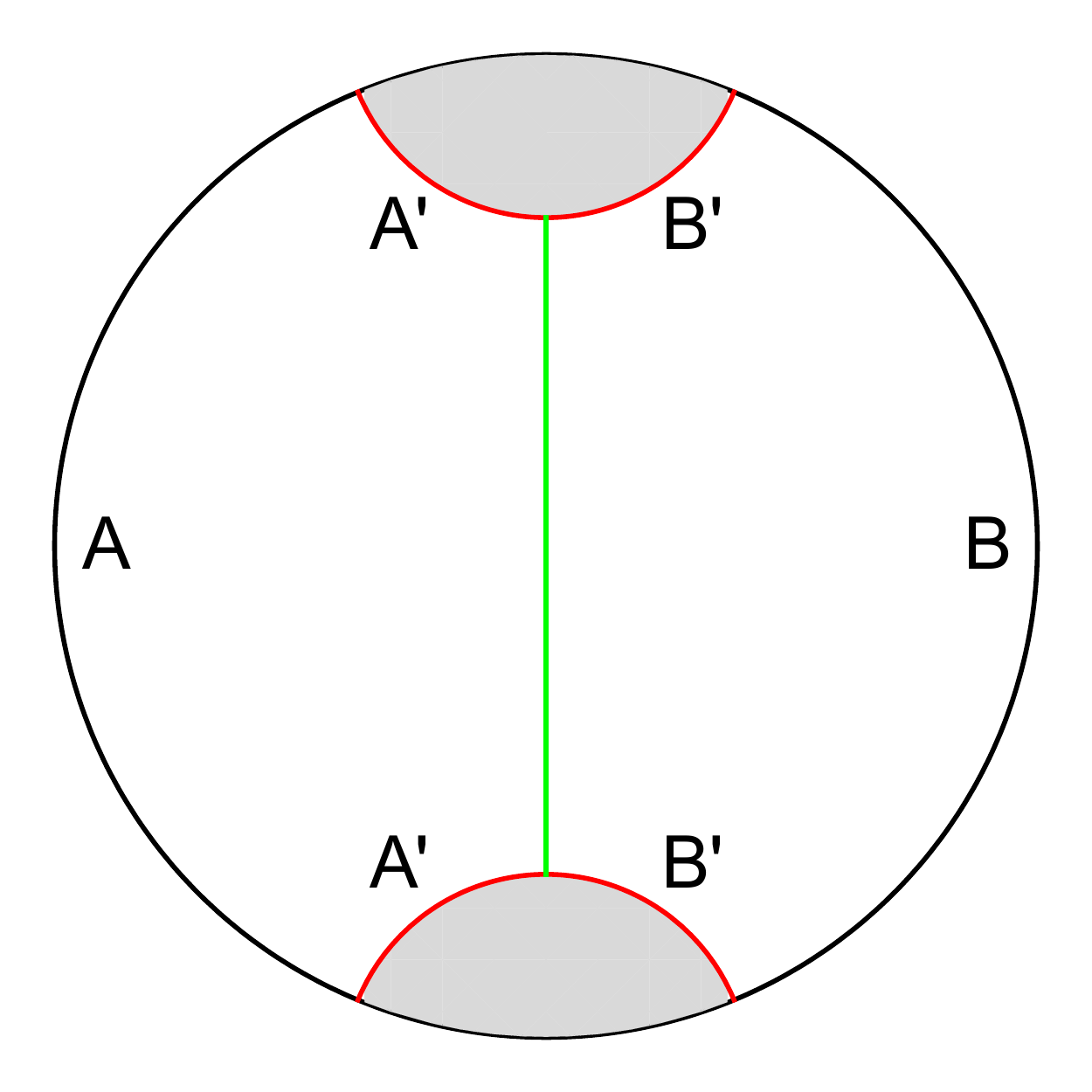}
    \includegraphics[width=4.5cm]{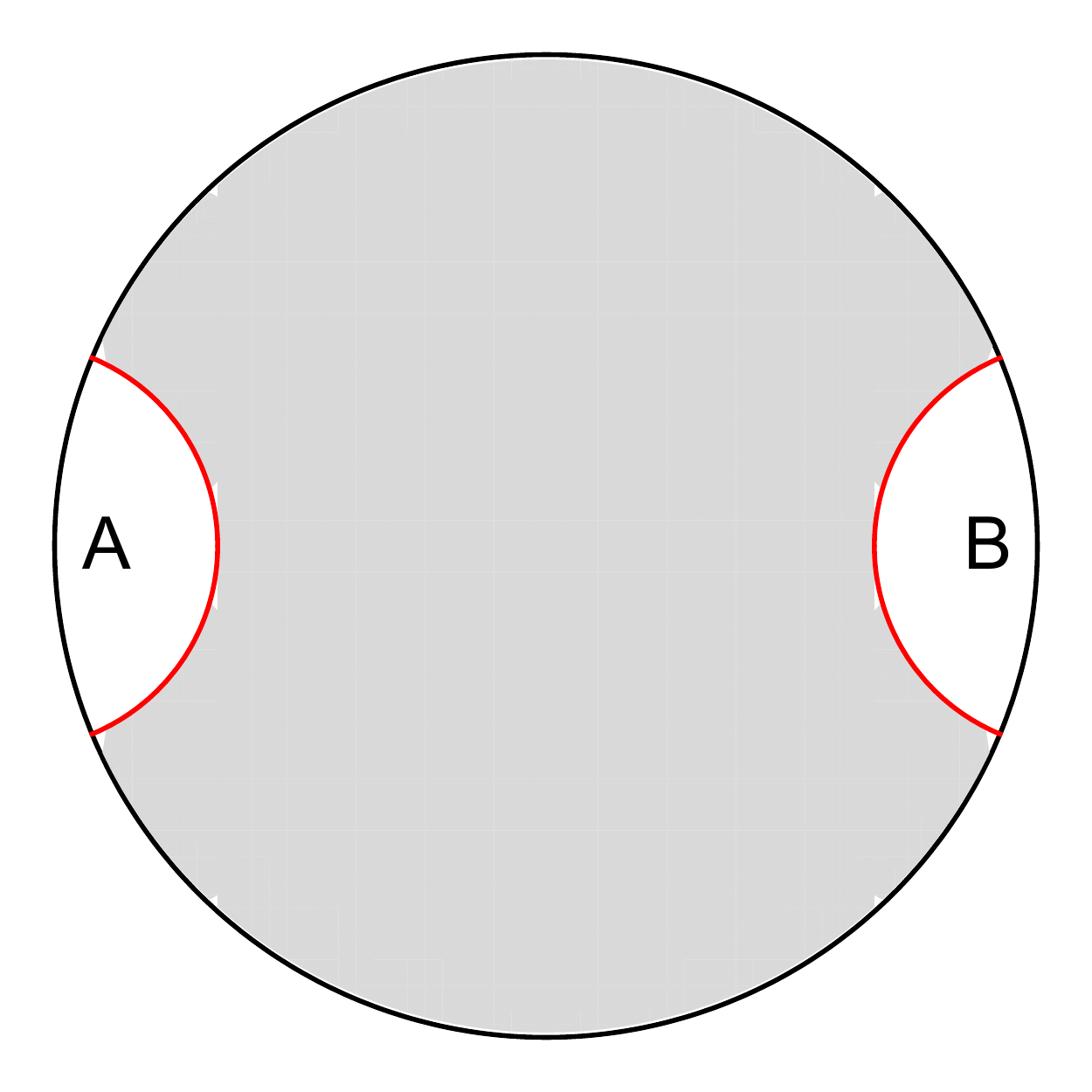}
    \includegraphics[width=4.5cm]{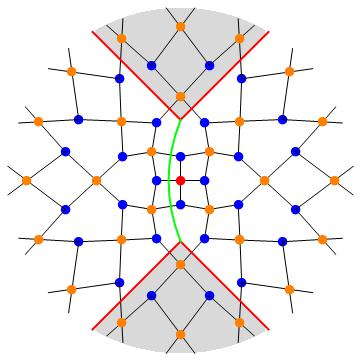}
    \caption{Left: For sufficiently large $A$ and $B$, the entanglement wedge is connected (the RT surface $\Sigma$ is shown in red) and the $E_{p}$ is computed by the length of the green geodesic $X$. Center: For small $A$ and $B$, the entanglement wedge is disconnected and the $E_{p}$ is zero. Right: Tensors under a causal cut in MERA (red line) can be gotten rid of by a unitary transformation. In each case, the region to be cut out is shaded in gray. }
    \label{fig:PlotEP}
\end{figure}

Suppose we have a geometry $M$ dual to some pure state $|\psi\rangle_{ABC}$. We want a holographic prescription for computing the entanglement of purification of the state on $AB$, which we will refer to as $E_{ph}{(A:B)}$. Our proposal is as follows. Let $\Sigma$ be the RT surface associated with the combined region $AB$. The spatial region bounded by $A$, $B$, and $\Sigma$ is called the entanglement wedge \footnote{We are actually talking about a spatial slice in the entanglement wedge. The entanglement wedge itself is the codimension-0 region in the bulk which is the bulk domain of dependence of any spacelike surface bounded by the HRT surface and the boundary region.}. We consider the entanglement wedge as a new holographic geometry with boundary $A \cup B \cup \Sigma$, i.e. by discarding all the geometry from $\Sigma$ to the old boundary $C$. The prescription is to find the minimum area surface $X$ which can end on $\Sigma$ that separates $A$ from $B$. The $E_{ph}$ is then given by:
\begin{equation}
    E_{ph}{(A:B)} = \frac{\mathrm{Area}{(X)}}{4G_{N}}
\end{equation}
where $G_{N}$ is Newton's constant. We illustrate this for two disjoint boundary intervals in global $AdS_{3}$ in Figure~\ref{fig:PlotEP}. As can be seen from this figure, the $E_{ph}$ in this case is only nonzero when the entanglement wedge is connected. In essence, $E_{ph}$ is the minimal cross-section of the entanglement wedge.\\
Note that, in the limit where $B$ is the complement of $A$ (in other words $\rho_{AB}$ is pure), our prescription reduces to the Ryu-Takayanagi formula. This is our first consistency check, since $E_{p}{(A:B)} = S{(A)} = S{(B)}$ for a pure state as argued previously.\\
This prescription has an alternative description. Imagine breaking $\Sigma$ into two pieces, $\tilde{A}$ and $\tilde{B}$. Group $\tilde{A}$ with $A$ and $\tilde{B}$ with $B$ and view the combined regions as two boundary regions. The whole system $A\tilde{A}B\tilde{B}$ is in a pure state. Now calculate the entanglement between $A\tilde{A}$ and $B\tilde{B}$ using the usual RT formula. Finally, minimize the resulting entropy over $\tilde{A}$ and $\tilde{B}$. The result is again the minimal area surface which can end on $\Sigma$ that separates $A$ from $B$. The minimal $\tilde{A}$ is taken to be $A'$ and similarly for the minimal $\tilde{B}$ ($A'$ and $B'$ are labelled on Figure \ref{fig:PlotEP}). This second formulation makes the physical intuition more clear. The idea is to simplify the geometry $M$ as much as possible by removing the geometry outside the entanglement wedge of $AB$. This is accomplished using some operation on $C$. The effect is to replace $C$ with $\Sigma$. We then break up $\Sigma$ into two pieces such that the combined entropy, as computed by the RT formula, is as small as possible. The above intuition suggests that, if we restrict to holographic purifications, then then entanglement of purification is given by our minimal surface prescription. The more non-trivial claim is that it suffices to restrict to such holographic purifications.\\

We note that the idea of thinking about $\Sigma$ as part of the new boundary is especially natural from the viewpoint of tensor networks and their connection to the AdS/CFT correspondence. For example, we show an analog of $\Sigma$ in a MERA network in Figure~\ref{fig:PlotEP}). Similar pictures can be drawn for networks of perfect tensors or random tensors \cite{Pastawski:2015qua,Hayden:2016cfa}. In the MERA example we can remove tensors from the shaded region by a unitary transformation acting on the complement of $AB$ thereby simplifying the geometry of the tensor network. For example, the number of boundary legs in the purification of $AB$ has gone from six to four by removing tensors below the lower red cut in Figure~\ref{fig:PlotEP}).

\subsection{Sample calculations of $E_{ph}$: pure $AdS_{3}$}
In this subsection, we provide explicit formulae for the $E_{ph}$ in empty $AdS_{3}$.

\paragraph{Non-adjacent intervals in $AdS_{3}$.} First, consider the case where $A$ and $B$ are 2 non-adjacent intervals in global $AdS_{3}$. In this case the RT surface comes in 2 different topologies depending on the size and separation of the 2 intervals as illustrated in Figure~\ref{fig:PlotEP}): either (1) one component of the RT surface connects the endpoints of $A$ and the other one connects the endpoints of $B$, or (2) each component connects one endpoint of $A$ with one endpoint of $B$.

In the first case, no curve in the bulk separates $A$ from $B$ and we say that the $E_{ph}$ is zero. One could argue for this value of $E_{ph}$ by invoking the mutual information. In this regime, $S{(AB)}=S{(A)}+S{(B)}$ and $I{(A:B)}=0$. This implies that, to leading order in $N$ (in the large-$N$ limit), the reduced density matrix is a product state $\rho_{AB} = \rho_{A} \otimes \rho_{B}$. It can be seen that the $E_{p}$ of a product state is always zero. Apparently, according to our picture, the subleading $1/N$ corrections do not affect the $E_{p}$.

Finding $E_{ph}$ in the second case involves finding the shortest distance between 2 geodesics in the hyperbolic plane. This is a nontrivial exercise in hyperbolic geometry, and we relegate the details to Appendix \ref{App:Distance} and simply quote the result here. If we parametrize the two subsystems by  $A = (\phi_{1}-\alpha_{1},\phi_{1}+\alpha_{1})$ and $B = (\phi_{2}-\alpha_{2},\phi_{2}+\alpha_{2})$, then the $E_{ph}$ between the two geodesics is given by:
\begin{equation}\label{EphNonAdjacent}
    E_{ph} = \frac{\lads}{4G_{N}} \log{\left( \frac{(\sqrt{\Delta}+\sqrt{2\sin{\alpha_{1}}\sin{\alpha_{2}}})^{2}}{\Delta-2\sin{\alpha_{1}}\sin{\alpha_{2}}} \right)}
\end{equation}
\begin{equation}
    \Delta = \cos{(\alpha_{1}-\alpha_{2})}-\cos{(\phi_{1}-\phi_{2})}
\end{equation}
The formula above applies of course whenever the entanglement wedge is connected. Note that the formula only depends on $\phi_{1}$, $\phi_{2}$ through their difference, reflecting the rotational symmetry. Alternatively, if we parametrize the boundary intervals by their endpoints as $A = (\theta_{1},\theta_{2})$ and $B=(\theta_{3},\theta_{4})$ the formula becomes:
\begin{equation}\label{EphEndpointsNonAdj}
    E_{ph} = \frac{\lads}{4G_{N}} \log{\left\{ \frac{\left[ \sqrt{\sin{((\theta_{1}-\theta_{3})/2)}\sin{((\theta_{2}-\theta_{4})/2)}}+\sqrt{\sin{((\theta_{2}-\theta_{1})/2)}\sin{((\theta_{4}-\theta_{3})/2)}} \right]^{2}}{\sin{((\theta_{2}-\theta_{3})/2)}\sin{((\theta_{1}-\theta_{4})/2)}} \right\}}
\end{equation}
Also, for the special case $\alpha_{1}=\alpha_{2} \equiv \alpha$, $\phi_{1}=\frac{\pi}{2}$, $\phi_{2} = \frac{3\pi}{2}$ (i.e. two geodesic of the same size diametrically opposite each other) the above reduces to:
\begin{equation}\label{DSymmetric}
    E_{ph}{(\alpha)} = \frac{\lads}{4G_{N}} \log{\left( \frac{1+\sin{\alpha}}{1-\sin{\alpha}} \right)}
\end{equation}
This is the situation depicted on the left panel of Figure \ref{fig:PlotEP}.

To get a sense of the formula (\ref{EphNonAdjacent}), we can vary one endpoint of one of the two geodesics (with the other 3 endpoints kept fixed) and plot the $E_{ph}$ as a function of the varying endpoint. This is what we show in Figure \ref{fig:Monotonicity} below. Note that the $E_{ph}$ is only nonzero in a certain range of the parameters.
\begin{figure}[H]
    \centering
     \includegraphics[width=7cm]{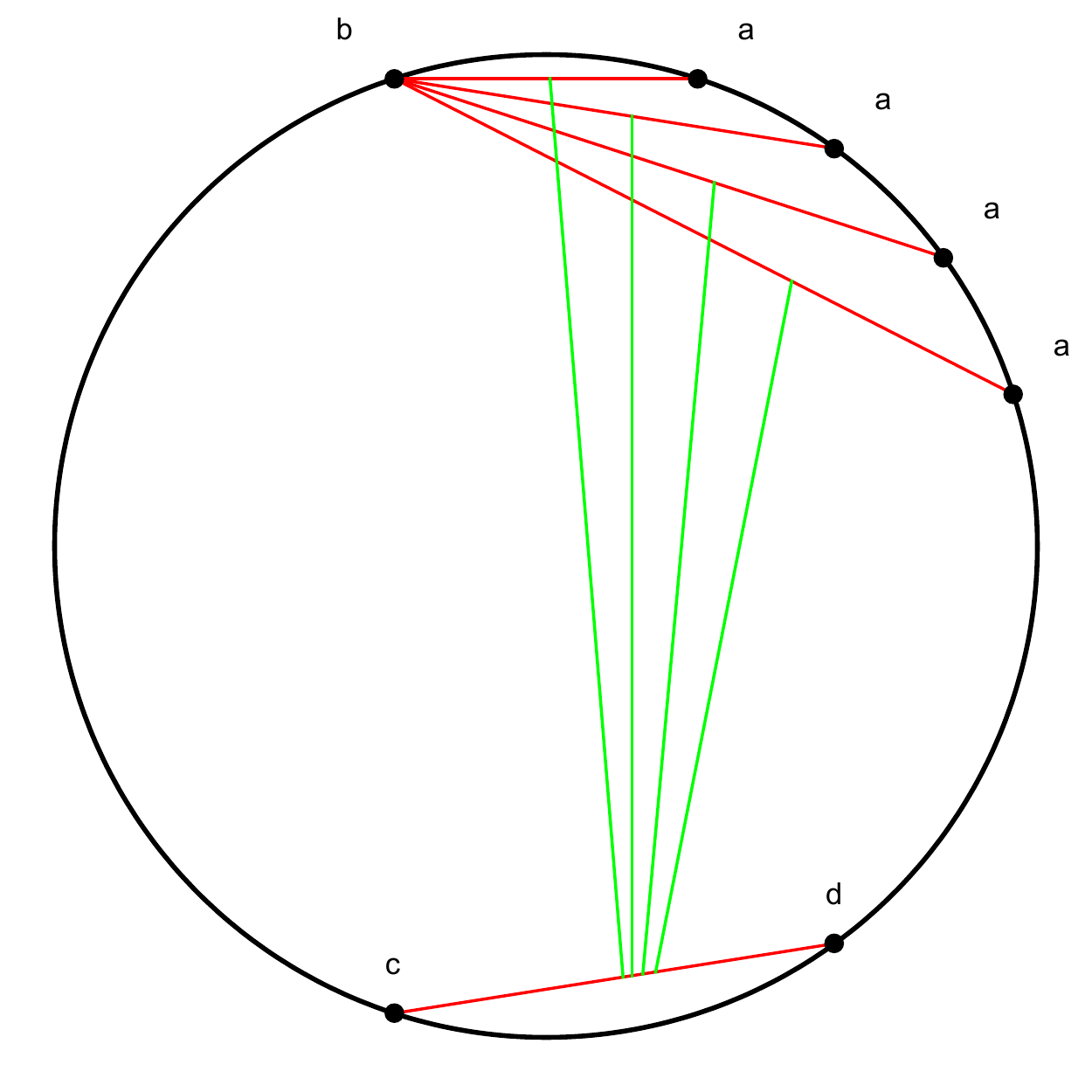}
    \includegraphics[width=8cm]{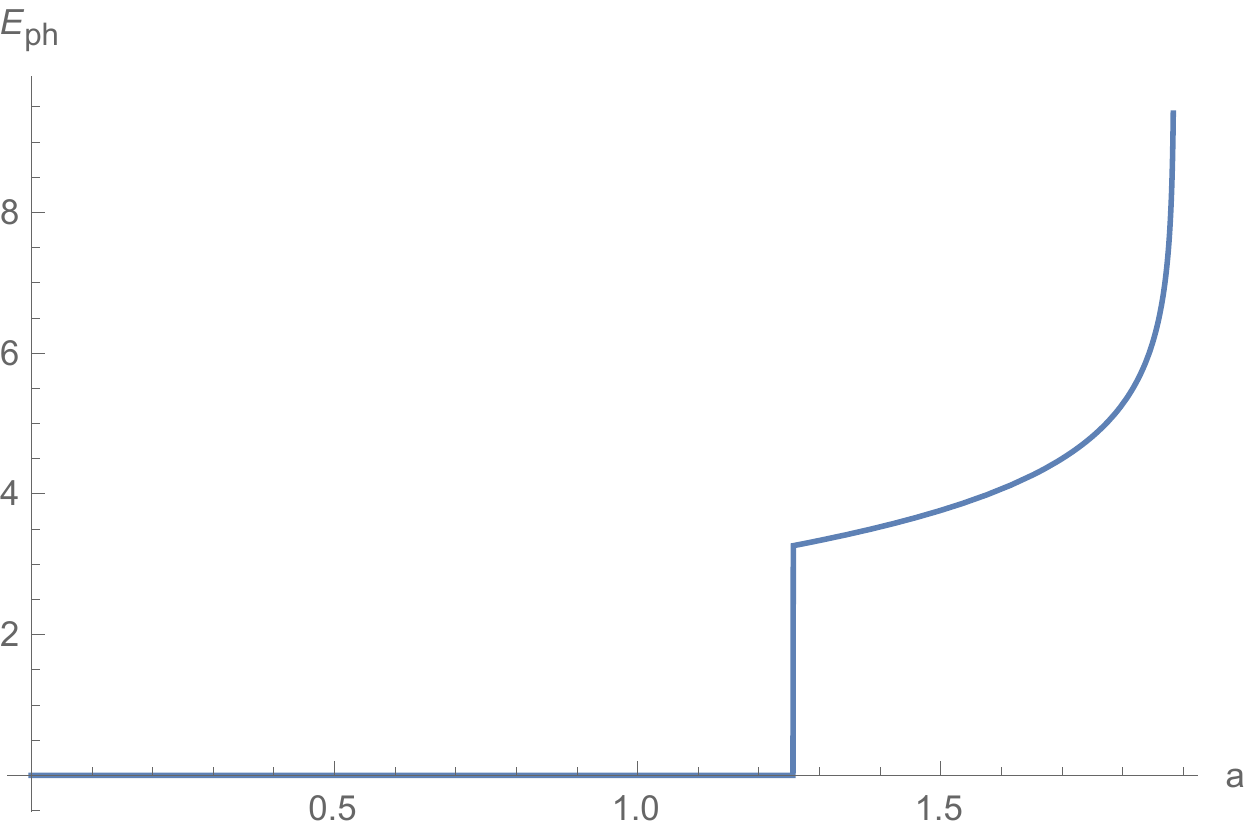}
    \caption{Left: we vary the position of $a$ and keep $b$, $c$, $d$ fixed. The values chosen here are $b=0.6\pi$, $c=1.4\pi$ and $d = 1.7\pi$. The green geodesics are the shortest curves connecting $(ab)$ to $(cd)$. Here we use the Beltrami-Klein coordinate system (explained in Appendix \ref{App:Distance}), in which geodesics are straight lines. Right: Plot of the $E_{ph}$ as a function of $a$, over the range $a \in [0,b]$. The $E_{ph}$ diverges when $a=b$, and undergoes a phase transition near $a \approx 1.256$ (where the RT surface changes topology). We set $4G_{N}=1$.}
    \label{fig:Monotonicity}
\end{figure}

\begin{figure}[H]
    \centering
     \includegraphics[width=9cm]{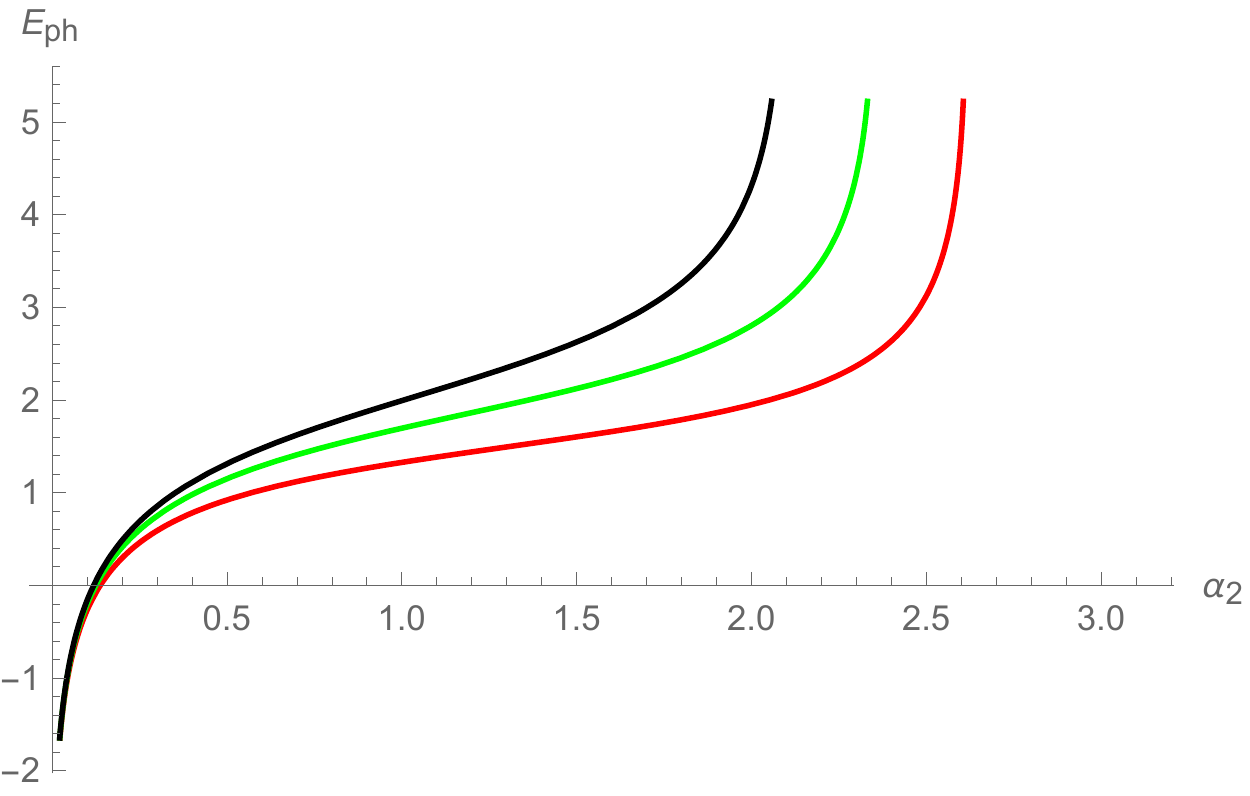}
    \caption{Plot of $E_{ph}$ for 2 adjacent intervals as a function of $\alpha_{2}$, at fixed $\alpha_{1}$. The values of $\alpha_{1}$ are: $\pi/6$ (red), $\pi/4$ (green) and $\pi/3$ (black). We set the cutoff $\epsilon$ to $0.1$ and $4G_{N}=1$.}
    \label{fig:EphPlotAdjacent}
\end{figure}

\paragraph{Adjacent intervals in AdS.}
Next we compute the $E_{ph}$ for two adjacent intervals, which is a special case of the non-adjacent case above, but we need to regulate the divergence. Consider 2 adjacent intervals $A$, $B$ on the boundary, with half-widths $\alpha_{1}$ and $\alpha_{2}$ respectively. The $E_{ph}$ in this case is the shortest distance from the common endpoint of $A$ and $B$ to the RT surface of $AB$, and it can be found using the same techniques as in the previous case of non-adjacent intervals. Note also that the $E_{ph}$ in this case is divergent whereas it is finite in the previous case. We relegate the details to Appendix \ref{App:Distance} again and only give the final result here:
\begin{equation}\label{EphAdjacent}
     E_{ph}{(\alpha_{1},\alpha_{2})} = \frac{\lads}{4G_{N}} \log{\left( \frac{2\sqrt{2}\csc{(\alpha_{1}+\alpha_{2})}\sin{\alpha_{1}\sin{\alpha_{2}}}}{\sqrt{\epsilon}}\right)} + \dots 
\end{equation}
where $\epsilon$ is a near-boundary cutoff (the geodesic is regulated at Beltrami-Klein radial coordinate $\lads (1-\epsilon)$), and $\dots$ stand for terms which vanish as $\epsilon \rightarrow 0$.\\
In particular, in the symmetrical case where the two adjacent intervals have the same half-width $\alpha_{1}=\alpha_{2} \equiv \alpha$, the above simplifies to \footnote{The cutoff $\epsilon$ can be converted to a cutoff in global radial coorinate $R_{c}$ by $R_{c} \approx \frac{\lads}{\sqrt{2\epsilon}}$}:
\begin{equation}\label{Ephsymmetric}
    E_{ph}{(\alpha)} = \frac{\lads}{4G_{N}} \log{\left( \sqrt{\frac{2}{\epsilon}} \tan{\alpha} \right)}
\end{equation}
We plot in Figure (\ref{fig:EphPlotAdjacent}) the $E_{ph}$ as a function of $\alpha_{2}$ for fixed values of $\alpha_{1}$. One can notice from the plot that the $E_{ph}$ is neither a convex nor a concave function of the boundary intervals' sizes. This is more or less expected, since the $E_{p}$ is known to be neither concave nor convex with respect to mixture of states \cite{2002JMP....43.4286T}.
Note that the $E_{ph}$ for adjacent intervals is essentially the mutual information (for the same choice of cutoff in the bulk, the two quantities differ by only $(\lads/4G_{N}) \log{2}$, see section \ref{Sec:HolographicIneq} for more details). Interestingly, the functional form of (\ref{EphAdjacent}) is also the same as that of the logarithmic negativity for 2 adjacent intervals in a CFT \cite{Calabrese:2012ew} (see also \cite{Chaturvedi:2016rft}).

\subsection{Sample calculations: 1-sided BTZ black hole}

Next, we present some sample calculations for the BTZ black hole. We focus on the 1-sided black hole in this subsection, with metric \cite{Banados:1992wn}:
\begin{equation}
    ds^{2} = -\frac{r^{2}-r_{+}^{2}}{\lads^{2}} dt^{2} + \frac{\lads^{2}}{r^{2}-r_{+}^{2}} dr^{2} + r^{2}d\phi^{2}
\end{equation}
and will consider the 2-sided black hole in the next subsection. The Hawking temperature is given by $\beta/\lads = 2\pi \lads /r_{+}$. We distinguish between 2 cases: (1) when the entanglement wedge is topologically trivial (i.e. connected and simply connected), and (2) when the entanglement wedge is not simply connected due to the inclusion of the horizon.\\
\paragraph{Case (1).} In the first case, we can use the fact that BTZ is a quotient of global AdS. Thus it is straightforward to map formulae (\ref{EphNonAdjacent}) and (\ref{EphAdjacent}) from AdS to derive the analogous formula for $E_{ph}$ in BTZ. We do not even need the full coordinate transformation from global $AdS$ to BTZ, but only the transformation of the boundary coordinates. It is known that the coordinate transformation from AdS to BTZ reduces to a conformal transformation on the boundary:
\begin{equation}
    \tan{\left[ \frac{1}{2} \left( \frac{\tau}{\lads} \pm \theta \right) \right]} = \tanh{\left[ \frac{r_{+}}{2\lads} \left( \frac{t}{\lads} \pm \phi \right) \right]}
\end{equation}
Here $(\tau,\theta)$ are the global AdS time and angle coordinates, and $(t,\phi)$ are the BTZ time and angle coordinates. In particular, on the slice $\tau=0$ (or equivalently $t=0$) we have:
\begin{equation}
    \tan{\left(\frac{\theta}{2}\right)} = \tanh{\left(\frac{r_{+}}{2\lads}\phi\right)}
\end{equation}
In particular, this implies:
\begin{equation}\label{AdSToBTZRelation}
    \sin{\left( \frac{\theta_{2}-\theta_{1}}{2} \right)} = \frac{\sinh{(\frac{r_{+}}{2 \lads} (\phi_{2}-\phi_{1}))}}{\sqrt{\cosh{(r_{+}\phi_{2}/\lads)}\cosh{(r_{+}\phi_{1}/\lads)}}}
\end{equation}
Next, we substitute the above into formula (\ref{EphEndpointsNonAdj}) for the $E_{ph}$ of two non-adjacent intervals in BTZ (such that the entanglement wedge is connected and simply connected):
    \begin{equation}\label{EphEndpointsNonAdjBTZ}
    E_{ph} = \frac{\lads}{4G_{N}} \log{\left\{ \frac{\left[ \sqrt{\sinh{(\frac{r_{+}}{2\lads}(\phi_{1}-\phi_{3}))}\sinh{(\frac{r_{+}}{2\lads}(\phi_{2}-\phi_{4}))}}+\sqrt{\sinh{(\frac{r_{+}}{2\lads}(\phi_{2}-\phi_{1}))}\sinh{(\frac{r_{+}}{2\lads}(\phi_{4}-\phi_{3}))}} \right]^{2}}{\sinh{(\frac{r_{+}}{2\lads}(\phi_{2}-\phi_{3}))}\sinh{(\frac{r_{+}}{2\lads}(\phi_{1}-\phi_{4}))}} \right\}}
\end{equation}
The case of two adjacent intervals in BTZ can be similarly handled.

\paragraph{Case (2).} Next, we discuss the more complicated case where the entanglement wedge has a hole due to the horizon. In this case, the surface computing the $E_{ph}$ becomes disconnected.

\begin{figure}[H]
    \centering
    \includegraphics[width=5cm]{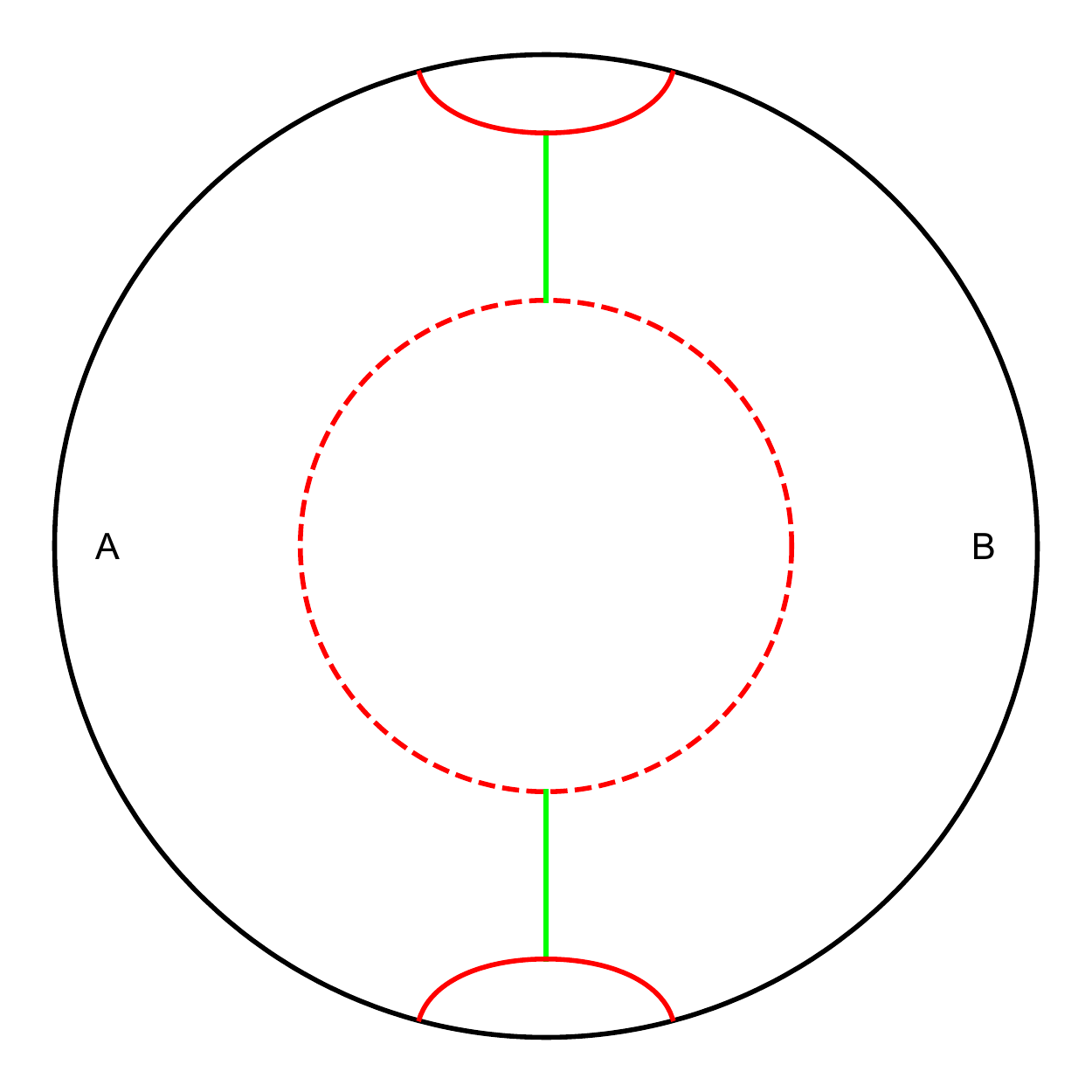}
    \includegraphics[width=5cm]{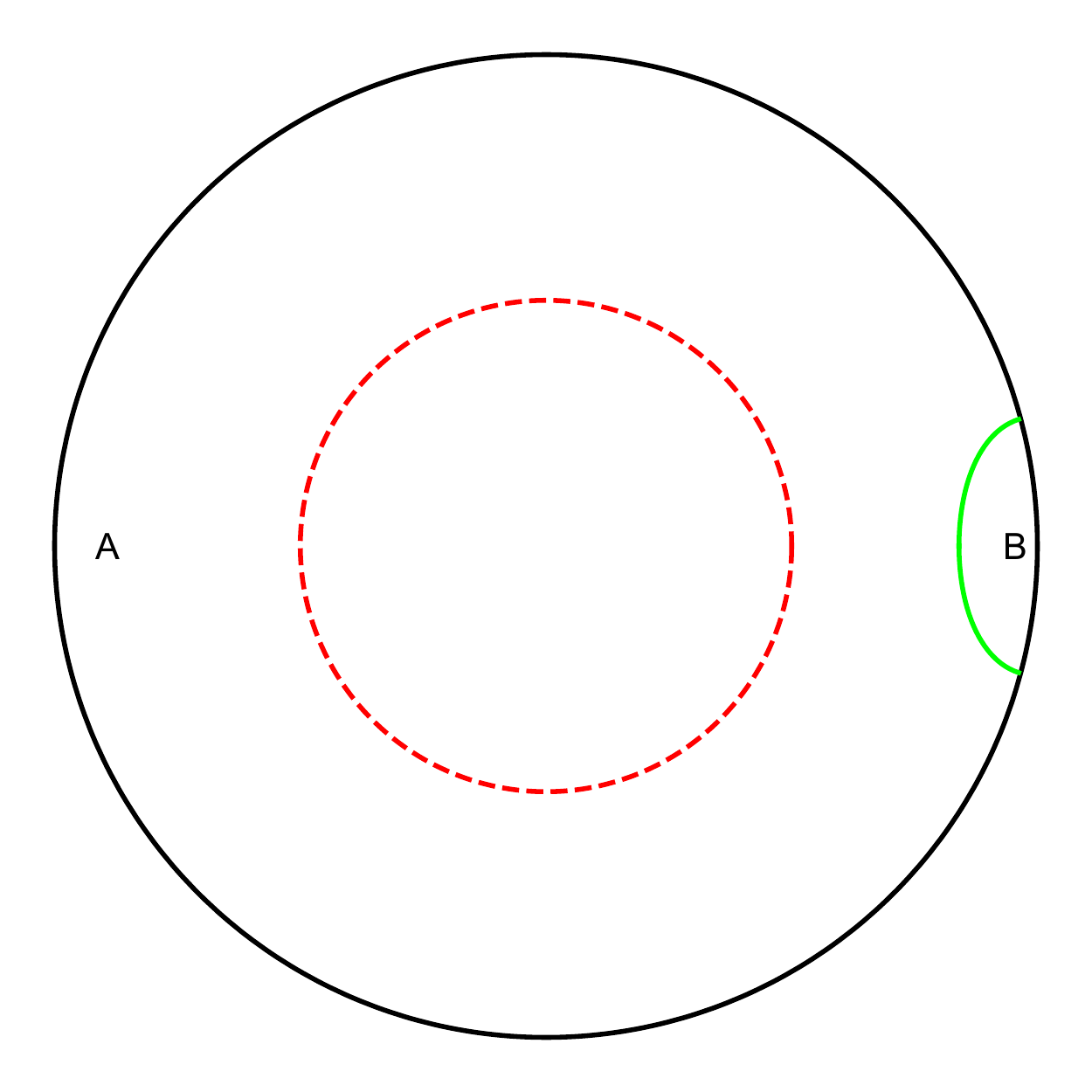}
    \caption{Left: The $E_{ph}$ geodesic is in green, and the RT surface (including the horizon) is in red. Right: When the Araki-Lieb inequality is saturated, the $E_{ph}$ coincides with $S(B)$.}
    \label{fig:BTZ}
\end{figure}

\begin{figure}[H]
    \centering
     \includegraphics[width=7cm]{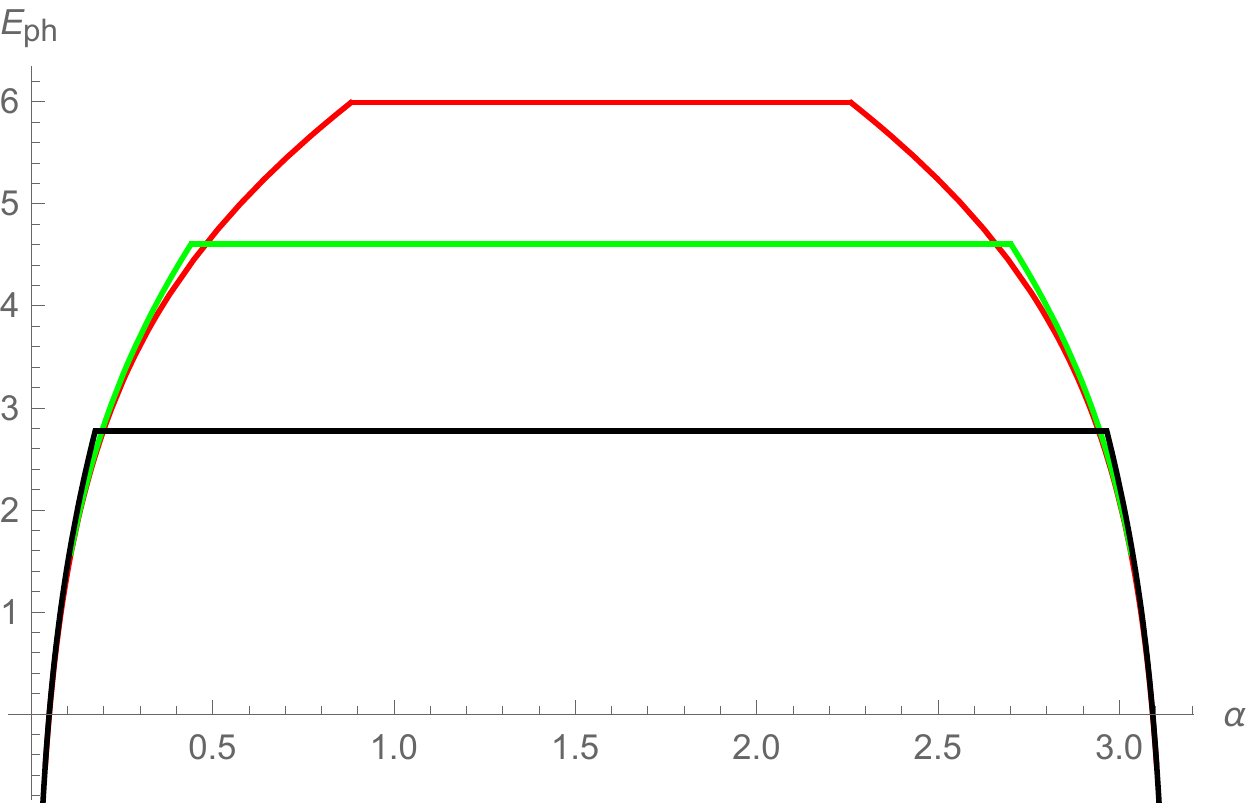}
     \includegraphics[width=7cm]{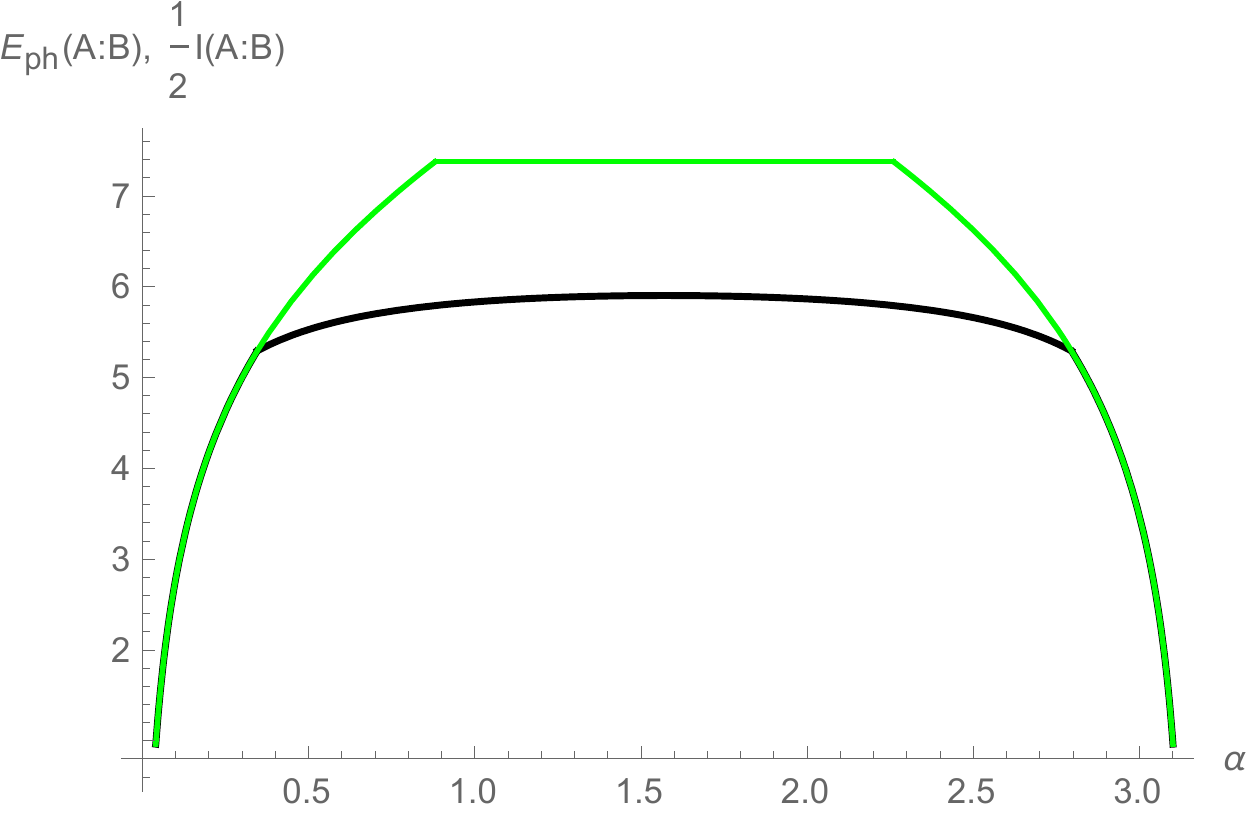}
    \caption{Left: Plot of $E_{ph}$ for the case where $A$ has half-width $\alpha$ and $B$ has half-width $\pi-\alpha$ for 3 different choices of the horizon: $r_{+}/L = 1$ (red), $r_{+}/\lads = 2$ (green) and $r_{+}/L = 5$ (black). Right: Plot of the $E_{ph}$ (green) and half the mutual information (black) as a function of $\alpha$, with $r_{+}/\lads =1$. In both panels, we set the radial cutoff to $r_{c}/\lads=10$ and $4G_{N}=1$.}
    \label{fig:EphPlotBTZ}
\end{figure}

Let us consider a few simple special cases, starting with the case where $A$ and $B$ are of the equal size, each slightly smaller than half the boundary circle (on one side of the BTZ black hole), as depicted in the left panel of Figure \ref{fig:BTZ}. Then the RT surface for $AB$ has 3 connected components, one of which is the horizon. The EP geodesic extends in the radial direction as depicted in Figure \ref{fig:BTZ}.
The $E_{ph}$ is:
\begin{equation}
E_{ph} = \frac{2}{4G_{N}} \int_{r_{+}}^{r_{*}} \frac{\lads}{\sqrt{r^{2}-r_{+}^{2}}} dr = \frac{\lads}{2G_{N}} \log{\left( \frac{r_{*}}{r_{+}} + \sqrt{\left(\frac{r_{*}}{r_{+}}\right)^{2}-1} \right)}
\end{equation}
where $r_{*}$ is radial coordinate of the deepest point of the RT components that go to the boundary. It is related to the half-width $\alpha$ of the boundary intervals $A$ or $B$ by:
\begin{equation}
    r_{*} = r_{+}\coth{\left(\frac{r_{+}}{\lads} \left( \frac{\pi}{2} - \alpha \right) \right)}
\end{equation}
In terms of $\alpha$, the $E_{ph}$ can be written as:
\begin{equation}
    E_{ph}{(r_{+},\alpha)} = \frac{\lads}{2G_{N}} \log{\left[ \coth{\left( \frac{r_{+}}{2\lads} \left( \frac{\pi}{2}-\alpha \right) \right)} \right]}
\end{equation}
In particular, when $\alpha = \frac{\pi}{2}$ the $E_{ph}$ is divergent. The regularized $E_{ph}$ in this case is:
\begin{equation}\label{Ephplateau}
    E_{ph}{(r_{+},\alpha=\frac{\pi}{2})} = \frac{\lads}{2G_{N}} \log{(r + \sqrt{r^{2}-r_{+}^{2}})} \bigg|_{r_{+}}^{r_{c}} =  \frac{\lads}{2G_{N}} \log{\left( \frac{2r_{c}}{r_{+}} \right)}
\end{equation}
Next, consider the case where the union of $A$ and $B$ is the whole boundary circle, say $A$ has half-width $
\alpha$ and $B$ has half-width $\pi-\alpha$. Moreover, suppose $\alpha$ is either sufficiently large or sufficiently small enough that we are in the ``entanglement plateau regime'' \cite{Hubeny:2013gta}. This means the Araki-Lieb inequality $S(AB) = |S(A) - S(B)|$ is saturated, which in turn implies that the $E_{p}$ coincides with the entanglement entropy of the smaller subsystem, and the $E_{ph}$ is computed by the RT surface for the smaller region. This is depicted in the right panel of Figure (\ref{fig:BTZ}).\\
Now let us vary $\alpha$ from $0$ to $\pi/2$. Initially $E_{ph} = S{(A)}$. Explicitly:
\begin{equation}\label{BTZGeodesicLength}
    E_{ph}{(\alpha,r_{+})} = \frac{\lads}{2G_{N}} \log{\left[ \frac{2r_{c}}{r_{+}}\sinh{\left(\frac{r_{+}}{\lads} \alpha \right)} \right]}
\end{equation}
At the critical angle  $\alpha_{crit,EP}$ given by:
\begin{equation}
    \alpha_{crit,EP}^{1} = \frac{\lads}{r_{+}} \mathrm{arcsinh}{(1)}
\end{equation}
the RT surface exchanges dominance with a new saddle: the two radial geodesics crossing the horizon as depicted on the left panel of (\ref{fig:BTZ}) and its $E_{ph}$ is given by (\ref{Ephplateau}). As $\alpha$ keeps increasing, the $E_{ph}$ levels off for a while since the surface remains two radial geodesics despite the change in $\alpha$. At the second critical angle:
\begin{equation}
    \alpha_{crit,EP}^{2} = \pi - \frac{\lads}{r_{+}} \mathrm{arcsinh}{(1)}
\end{equation}
the $E_{ph}$ surface snaps back to being the RT surface again. We plot the $E_{ph}$ versus $\alpha$ for 3 different choices of the horizon (or temperature) on the left panel of Figure (\ref{fig:EphPlotBTZ}), and we plot both the $E_{ph}$ and half the mutual information for a choice of $r_{h}$ on the right.
Let us now elaborate on Figure~\ref{fig:EphPlotBTZ}. The fact that the $E_{ph}$ levels off for $\alpha$ close enough to $\pi/2$ can be accounted for by the fact that correlations in a thermal state are short-range (they are cut off at the thermal scale). Note that the mutual information, like the $E_{p}$, is also a measure of the total correlation in the quantum state, and therefore should be expected to saturate for larger values of $\alpha$. This is indeed the case as can be seen from Figure (\ref{fig:EphPlotBTZ}). Interestingly, the mutual information saturates at an angle somewhat smaller than the angle of $E_{ph}$ saturation. That this happens is a consistency check for our proposal: it implies that whenever the Araki-Lieb inequality is saturated, then $E_{ph}$ is indeed given by the entanglement entropy of the smaller region.

\begin{figure}[H]
    \centering
     \includegraphics[width=10cm]{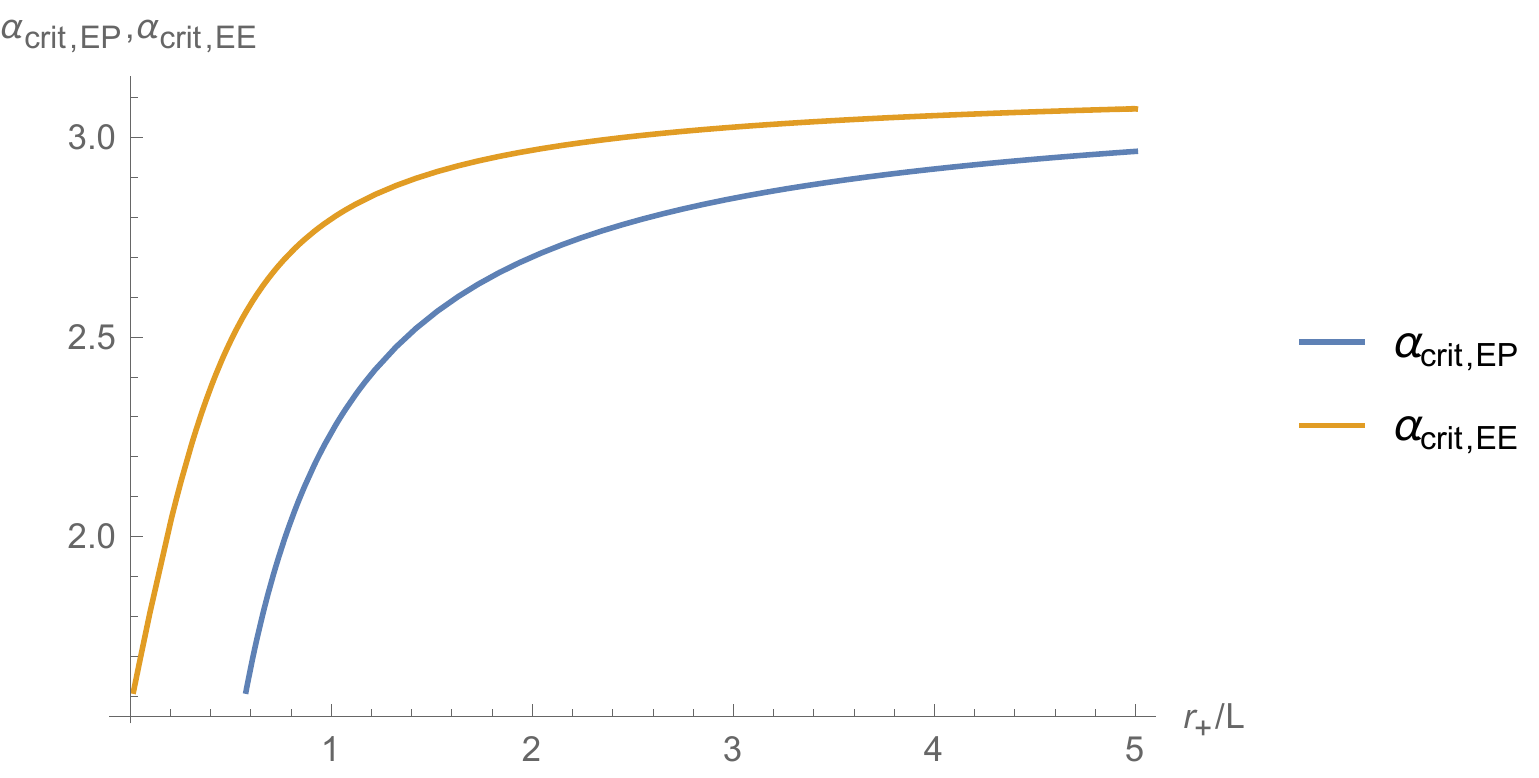}
    \caption{The two critical angles $\alpha_{crit,EE}$ and $\alpha_{crit,EP}$ versus $r_{+}/\lads$. Only the range $r_{+}/\lads > 1$ is physically relevant.}
    \label{fig:CritAngles}
\end{figure}

On the right panel of Figure (\ref{fig:EphPlotBTZ}), we have picked a particular value for the horizon. It is interesting to compare the two critical angles $\alpha_{crit,EP}^{2}$ and $\alpha_{crit,EE}$ as a function of the horizon. If $\alpha_{crit,EP}^{2} > \alpha_{crit,EE}$ for some horizon size, then the argument above regarding the Araki-Lieb inequality would be in trouble! Recall that $\alpha_{crit,EE}$ is given by:
\begin{equation}
    \alpha_{crit,EE} = \frac{\lads}{r_{+}} \mathrm{arccoth}{\left[2\coth{\left( \frac{\pi r_{+}}{\lads} \right)} - 1 \right]}
\end{equation}
We plot in Figure (\ref{fig:CritAngles}) the two critical angles as a function of $r_{+}/\lads$. As can be seen from the plot, we always have $\alpha_{crit,EP} < \alpha_{crit,EE}$ and we do have a consistent picture (i.e. $E_{ph} = S{(B)}$ whenever Araki-Lieb is saturated).

\section{Numerical calculation of $E_p$  via finite-temperature matrix product state algorithms}\label{sec:numerics}

Calculating $E_p$ exactly requires a global minimization over the space of purifications --- a problem that is numerically difficult even for small wavefunctions.
The existence of a geometric interpretation of $E_p$, however, suggests that locality can be exploited during the minimization process.
For numerical purposes, the locality of a many body state can be captured using a \emph{tensor network} ansatz. Here we explain how $E_p$ can be approximately computed in 1D using such methods. In fact, as discussed by Hauschild, et al. \cite{Hauschild}, the solution suggests a potentially dramatic speedup of finite-temperature DMRG calculations which should prove useful in its own right.

	In 1D, zero-temperature tensor network algorithms such as DMRG rely on the representation of a pure state as a matrix product state (MPS).\cite{ fannes, ostlund} MPSs are a class of variational ansatz  defined by the property that the entanglement entropy for a bipartition of the state into left and right regions is bounded from above by $S_{L:R} \leq \log(\chi)$.
Here $\chi$ is the ``bond-dimension'' of the MPS  - more entanglement can be captured by using larger $\chi$, but the computational cost generally scales as $\chi^3$.

\begin{figure}[t]
    \centering
    \includegraphics[width=0.7\textwidth]{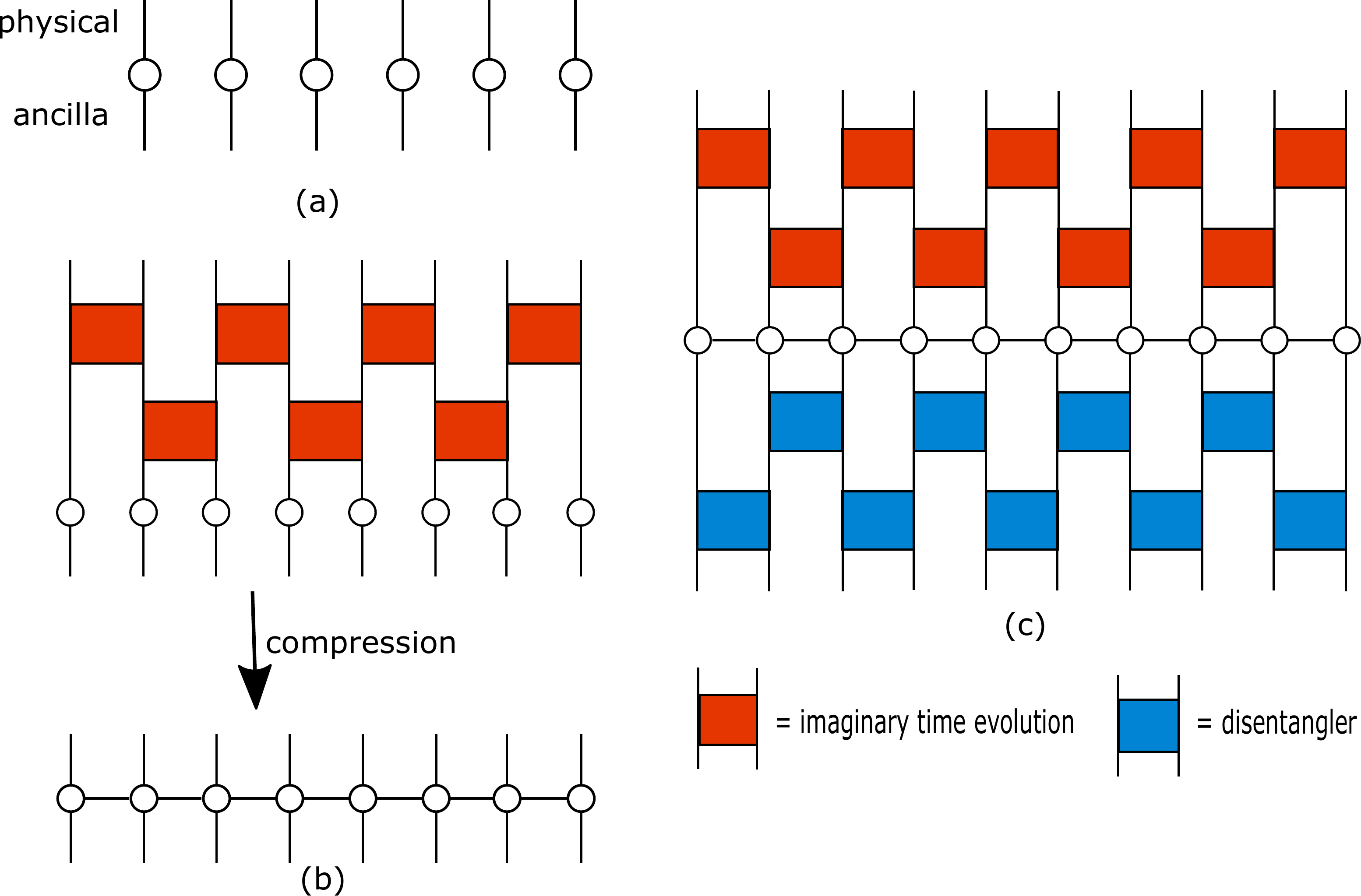}
    \caption{
        (a) The initial $|TFD,\beta=0\rangle$ MPS is a trivially entangled state.
        (b) After the application of the time evolution operator (red boxes, Trotter decomposed on to even an dodd bonds) on to the physical legs, the MPS is compressed following the usual TEBD algorithm as and results in an MPS with entanglement.
        After this step, we perform the disentangling sweep as described in the text.
        (c) The final form of tensor network produced by our algorithm after a single iteration.
}%
\label{fig:mps}
\end{figure}

When numerically simulating a mixed state $\hat{\rho}$, one can either represent $\hat{\rho}$ as a matrix product operator (MPO),\cite{verstraetetebd} or instead purify $\hat{\rho}$ and represent the purification as a MPS.\cite{schollwock}
Purifications have several advantages over density operators; for instance the density matrix will remain positive definite by construction, regardless of numerical errors. However, as discussed there is a large space of possible purifications, and the choice may drastically effect the numerical difficulty.\cite{Karrasch2013}
For equilibrium calculations, it is standard to use the  ``thermofield double'' (TFD) purification,
\begin{equation}
    \ket{ \tfd, \beta} = \frac{1}{\sqrt{Z(\beta)}}\sum_{n} e^{-\beta E_n/2} \ket{n}\ket{\tilde{n}}
    \label{eq:tfd}
\end{equation}
where $\left| n \right\rangle$,$\left|\tilde{n}\right\rangle$ are the $n$th eigenstate of $\mathcal{H}$ with energy $E_n$, on the physical and ancilla degrees of freedom respectively, $\beta$ is the inverse temperature, and $Z(\beta)=\sum_n e^{-\beta E_n}$ is the partition function.
In this case the Hilbert space of the ancilla is identical to the physical one, so locality can be preserved by doubling each degree of freedom  in the 1D chain. The MPS ansatz for the TFD state thus looks like a ``caterpillar'' (Figure~\ref{fig:mps}), just like the MPO representation of $\hat{\rho}$ would, but the prescription for calculating observables differs.

	The MPS representation of the TFD state is straightforward to obtain, for instance using the time-evolving block decimation (TEBD) algorithm.\cite{VidalTEBD,verstraetetebd,feiguin, schollwock}
At infinite temperature, $\beta=0$, the TFD state can be constructed by preparing each physical degrees of freedom into a maximally entangled state with its corresponding ancilla, e.g., for a spin-$1/2$ chain we have
\begin{equation}
    |\text{TFD}, \beta = 0\rangle = 2^{-L/2}\prod_j(|\uparrow\tilde{\uparrow}\rangle_{j} + |\downarrow\tilde{\downarrow}\rangle_{j})
    \label{eq:inftemptfd}
\end{equation}
where $|\cdot\tilde{\cdot}\rangle_{j}$ denote the states of the physical and ancilla degrees of freedom on site $j$.
This has zero entanglement across any cut and can therefore be represented  by an MPS with bond dimension $\chi=1$.
To prepare a state at finite $\beta$ using TEBD,\cite{VidalTEBD, schollwock} we apply $e^{-\beta H/2}$ to the physical degrees of freedom by Trotterizing the imaginary time evolution into small local gates. During the application of the gates to the MPS, the entanglement of the TFD state grows, and hence the bond dimension $\chi$.

	Starting from the TFD purification, we may obtain other purifications by acting with a unitary $U_{\text{anc}}$ on the ancilla.
Since the difficulty of MPS calculations increases  with $\chi \sim e^S$, we can try and use this freedom to reduce the entanglement of the purification.\cite{Karrasch2013}
Clearly the TFD is not itself optimal; as $\beta \to 0$, the TFD puts both the physical and ancilla degrees of freedom into the ground state, $\ket{\tfd, \infty} = \ket{0} \ket{\tilde{0}}$, with entanglement  \emph{twice} that of the ground state. Very crudely speaking, this requires a bond dimension which is the square of the ground state's $\chi_{\tfd} \sim \chi^2_{\textrm{gs}}$. The optimal purification would instead put the ancilla into a product state, e.g. $\ket{0} \ket{\tilde{\uparrow}, \tilde{\uparrow} \cdots}$, which  requires only $\chi_{\textrm{gs}}$,  suggesting something approaching a quadratic speedup of finite temperature calculations might be possible.
Minimizing the entanglement of the  purification, and hence hopefully the $\chi$ of the MPS, is precisely the problem of calculating the entanglement of purification.
	
Of course, all of this relies on the ability to correctly find the optimizing unitary $U_{\text{anc}}$.
Given the TFD MPS, how do we best find the optimal unitary that minimizes entanglement entropy across a cut?
Moreover, minimizing entanglement across a single cut is not very useful, since a priori this may increase the entanglement across other cuts, so we really want to minimize the \emph{sum} of the entanglement entropy at each cut.
This is, of course, a very difficult problem that we do not have an exact solution to.

Nevertheless, we can attempt to find an approximate solution by appealing to locality and restricting the structure of $U_{\text{anc}}$ to a unitary circuit formed from the successive  application of local (here two-site) gates. We accomplish this practically as follows.\cite{Hauschild} Starting from the $\beta = 0$ TFD state, we apply a small time step of imaginary time evolution to the physical degree of freedom,  $e^{- \Delta \beta H / 2} \ket{\beta = 0}$, compressing the result as an MPS. We then act with a \emph{disentangling} unitary $U_{\text{anc}}(0)$ which acts only on the ancilla. The disentangler takes the form of a depth-two unitary circuit acting first on even, then on odd bonds, $U_{\anc}(0) =  \prod_{j\in \textrm{odd}} U^{[j, j+1]}_{\anc} \prod_{j\in \textrm{even}} U^{[j, j+1]}_\anc $.
Each $U^{[j, j+1]}_{\anc}$ only affects the entanglement  of the corresponding bond, so we may \emph{locally} (gate-by-gate) solve the minimization problem
\begin{align}
\tilde{E}_p = \min_{U^{[j, j+1]}_{\anc}} S_{\dots j: j+1 \cdots } \left( U_{\text{anc}}(0)  e^{- \Delta \beta H/2} \ket{\beta = 0} \right ), 	
\end{align}
first calculating the even-bond unitaries, and then calculating the odd-bond unitaries holding the former fixed. Numerical algorithms for minimizing entanglement over a local gate have been discussed elsewhere. \cite{vidalmera}
Other disentangling criteria are also possible - in this work we actually minimize the 2nd Renyi entropy for numerical efficiency (see Appendix~\ref{app:s2min}).
This defines the optimal $U^{[j, j+1]}_{\anc}$ to apply, and $\tilde{E}_p$ is defined from the minimum.
The  purification at the next step is then defined by $\ket{\Delta \beta} =  U_{\anc}(0) e^{- \Delta \beta H/2} \ket{\beta = 0}$.
We then continue the similarly, alternating application of $e^{- \Delta \beta H / 2}$ on the physical degrees of freedom with a layer of unitary disentangling $U_{\anc}(\beta)$ on the ancilla.
This builds up a state of the form shown in Figure~\ref{fig:mps}, where $U_\anc = \cdots U_\anc(2 \Delta \beta) U_\anc(\Delta \beta) U_\anc(0)$.

	A priori, the resulting purification need not be the optimal one, first because $U_\anc$ was restricted to the form of a unitary circuit, and second because we determined the value of the initial layers using the low-$\beta$ purification, \emph{independent} of the subsequent layers. Indeed, $\tilde{E}_p$ is rather noisy at  intermediate temperature, presumably an artifact of our algorithm.
Nevertheless, the numerical experiments reveal that the entanglement $\tilde{E}_p$ of the purification we obtain is remarkably consistent with the expected properties of the true entanglement of purification $E_p$, as we now explore.

We study the standard transverse field Ising model (TFIM) at its critical point,
\begin{equation}
    \mathcal{H}_\text{TFIM} = J\sum_{i}\sigma^z_i\sigma^z_{i+1} + h\sum_{i}\sigma^x_i
    \label{}
\end{equation}
with $J=h=1/2$, where $\sigma^x$,$\sigma^z$ are Pauli matrices. While this model is equivalent to a free fermion problem, we have verified that the results are insensitive to an integrability-breaking perturbation which is tuned to stay at the critical point.
We obtain the entanglement entropy as a function of subsystem size $L_A$, inverse temperature $\beta$, and total system size $L$, using the method just  discussed, which we will refer to as the disentangled entanglement entropy $\tilde{E}_p(L_A,\beta,L)$.
If our disentangling unitary were optimal, then $\tilde{E}_p$ would coincide with the entanglement of purification $E_p$.

\begin{figure}[h]
    \centering
    \begin{subfigure}{0.45\textwidth}
        \includegraphics[width=1.0\linewidth]{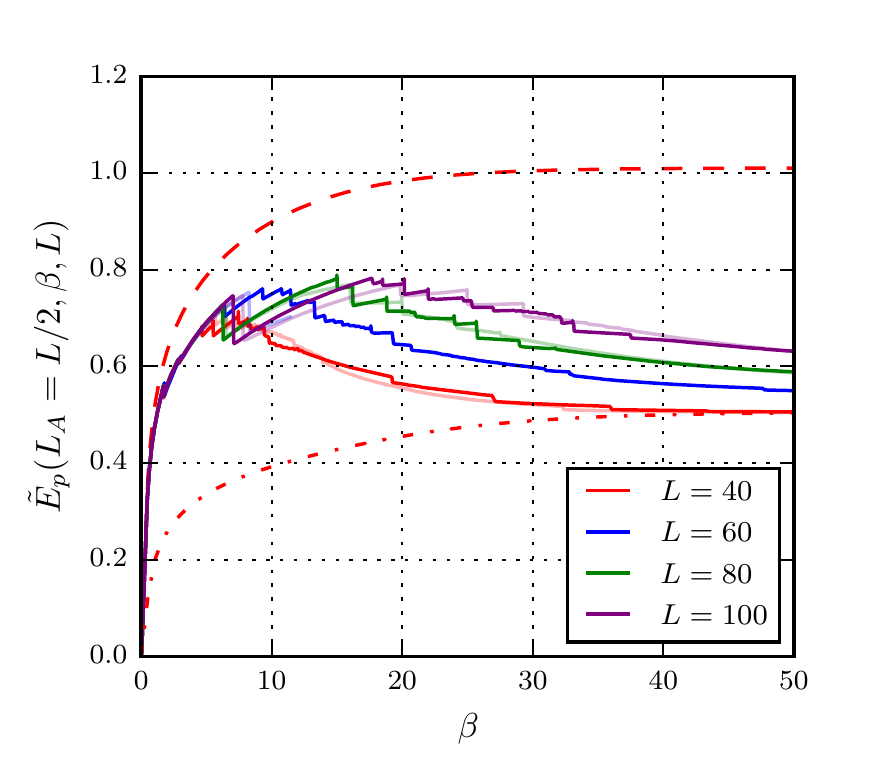}
    \end{subfigure}%
    \begin{subfigure}{0.45\textwidth}
        \includegraphics[width=1.0\linewidth]{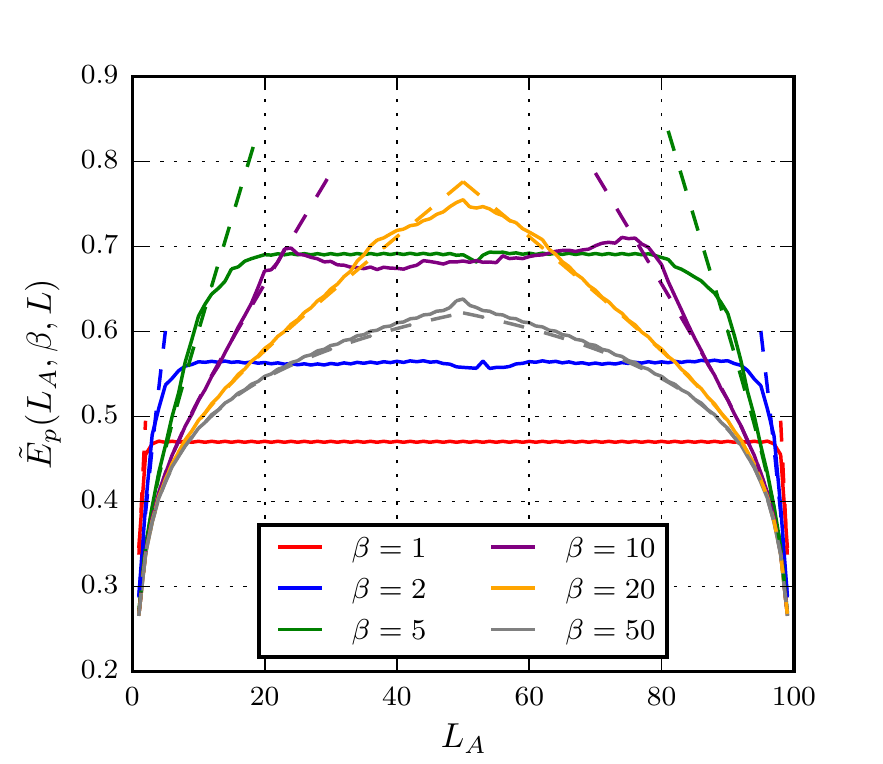}
    \end{subfigure}
        \caption{  \label{fig:tebd} \textbf{a}) The disentangled entropy $\tilde{E}_p(L_A,\beta,L)$ at the middle cut calculated using our disentangling algorithm ($L_A = L/2$).
        The calculation was done using a DMRG truncation error cutoff $\epsilon=10^{-14}$ and maximal bond dimension $\chi=48$.
        Faded lines correspond to calculations using $\chi=12$, which do not show a significant difference beyond the fluctuation from the disentangling.
        The dashed lines are results for the TFD state without disentangling and dash-dotted lines are half the mutual information to serve as an upper and lower bound respectively for $L=40$. \textbf{b}) Dependence of $\tilde{E}_p(L_A,\beta,L)$ on the subsystem size $L_A$ .
        Results for $L=100$ are shown with solid lines.
    Dashed lines show the minimum thermodynamic entropy $\min\{S(A),S(\bar{A})\}$ of the two subsystems subsystem, which matches excellently with $\tilde{E}_p$ up until saturation, as predicted from the holographic prescription.
        }%
\end{figure}
\begin{figure}[h]
    \centering
    \includegraphics[width=0.5\textwidth]{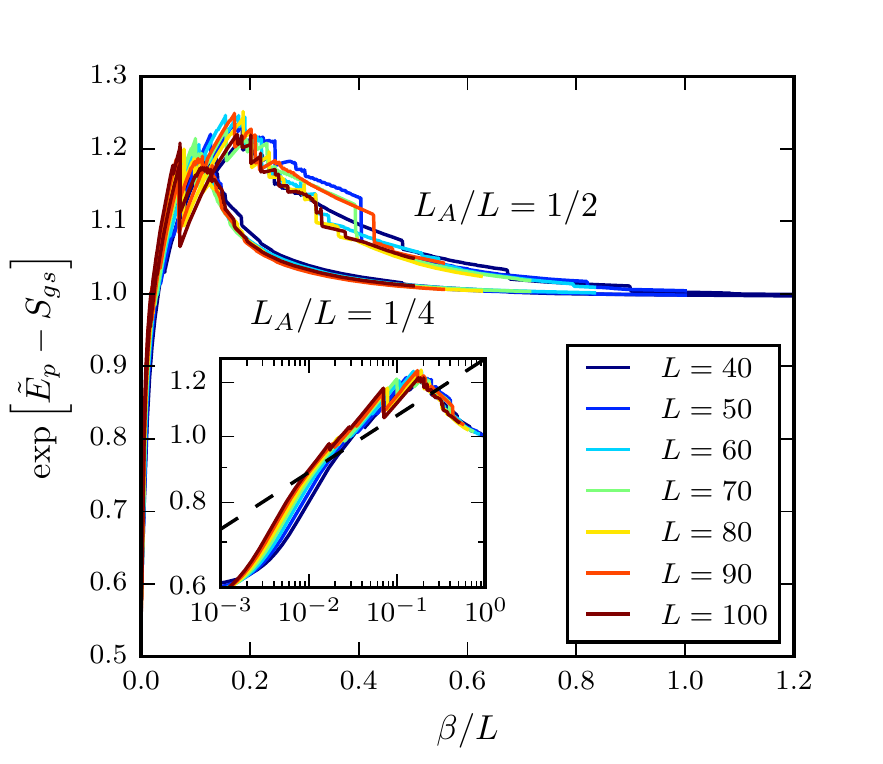}%
    \includegraphics[width=0.4\textwidth]{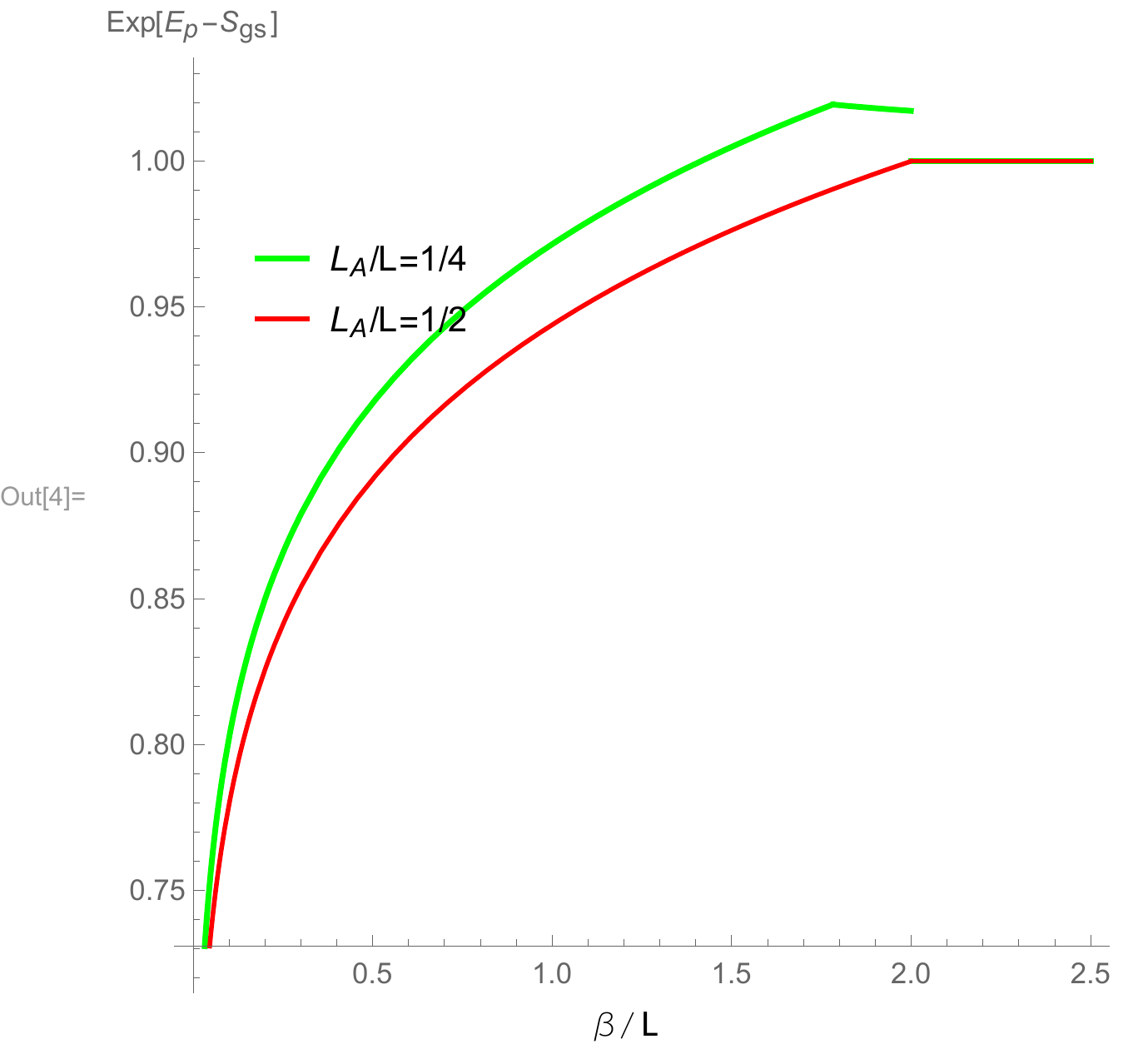}
    \caption{Left: The scaling form of the entanglement $e^{\tilde{E}_p - E_{gs}}=\tilde{f}(L_A/L,\beta/L)$ for $L_A/L=1/4$ and $1/2$, showing a collapse across different $L$. At $\beta \to \infty$, this approaches unity, consistent with $U_{\anc}$ completely disentangling the ancilla.
    The inset shows the same data on a log-log scale for $L_A/L=1/2$, and the dashed black line shows a $c/6$ power-law slope (from Eq~\ref{eq:bcftscaling}).
    Right: the scaling form from AdS/BCFT, with $c=1/2$ (the Ising value). The Hawking-Page transition occurs at $\beta/L = 2$. For very high temperatures, the $E_{p}$ surface drops vertically into the bulk. As $\beta/L$ increases, this surface can either exchange dominance with the one terminating on $Q$ before the Hawking-Page transition (as is the case for $L_{A}/L = 1/4$) or not (the case $L_{A}/L = 1/2$). }%
\label{fig:scalingform}
\end{figure}

In Figure~\ref{fig:tebd}, we show raw data for $\tilde{E}_p$ across the central cut ($L_A=L/2$) as a function of $\beta$ for a few system sizes $L$.
For reference, we also show the entanglement of the TFD state as an upper bound (obtained by TEBD \emph{without} disentangling) and half the mutual information as a lower bound (obtained via a thermal correlation matrix method~\cite{peschel}).
We believe the noise is due to a landscape of local minima in the entanglement minimization step (see Appendix~\ref{app:s2min}).
$\tilde{E}_p$ increases up to a maximum, before decreasing again and saturating the lower bound at high $\beta$.
Note that the saturation of the lower bound at $\beta \to \infty$ indicates that $U_{\anc}$ has successfully transformed the ground state of the ancilla $\ket{\tilde{0}}$ to an unentangled state, realizing the desired reduction $\chi_{\tfd} = \chi_{\textrm{gs}}^2 \to \chi_{\textrm{gs}}$ of the MPS.

Next, we examine the dependence of $\tilde{E}_p$ on the subsystem size $L_A$, shown in Figure~\ref{fig:tebd}b).
Also shown is the \emph{thermodynamic} von Neumann entropy $S_A, S_{\bar{A}}$ for the subsystem $A$ and its complement.
There are three clear regimes in the behavior of $\tilde{E}_p$: for small $L_A$, $\tilde{E}_p$ coincides with $S_A$, until it hits a plateau and saturates over a range of $L_A$.
Finally, as $L_A$ becomes the majority of the system, $\tilde{E}_p$ again coincides with the entropy of the smaller complement $S_{\bar{A}}$.

Remarkably, we find that $\tilde{E}_p$ satisfies the scaling form
\begin{align}
    e^{\tilde{E}_p(L_A,\beta,L)} = L^{c/6}f(L_A/L,\beta/L)
\label{eq:scaling_form}
\end{align}
where $c = \frac{1}{2}$ is the central charge, and $f$ is a universal function.
More conveniently, as we will show in Section~\ref{sec:bcft}, this can be expressed as $e^{\tilde{E}_p - S_{gs}}=\tilde{f}(L_A/L,\beta/L)$ becoming a universal function of $L_A/L$ and $\beta/L$, where $S_{gs}$ is the ground state entropy ($\tilde{f}$ is related to $f$ by a constant factor).
This is shown for $L_A/ L = 1/2, 1/4$ in  Figure~\ref{fig:scalingform}.

The qualitative agreement between the holographic and numerical results for the entanglement of purification is encouraging for both sides.
It is evidence that the holographic prescription $E_{ph}$ does indeed correspond to the entanglement of purification.
At the same time, another message is that although calculating $E_p$ numerically is difficult, it is possible to calculate it approximately with a practical algorithm. This result is also encouraging for numerical calculations of this type in general, where bond dimension is the limiting factor.
In our current algorithm, the computational gain from decreasing bond dimension is overshadowed by the cost of performing the disentangling at every time step, since our goal was to get as small an entanglement as possible.
In principle, the algorithm can be modified to include the disentangling step more sporadically (every few time steps), or only when necessary (if bond dimension goes above a certain value).

\subsection{Comparison with Holographic BCFT}\label{sec:bcft}

Here we compare the numerical results, which were obtained from a spin chain with open boundary conditions, to the holographic proposal in the case of open boundary conditions. Since the conformal field theory has open boundary conditions, the appropriate tool is now ``boundary conformal field theory" (BCFT), not to be confused with the conformal field theory at the boundary of AdS. The holographic calculations are based on an unproven but plausible proposal \cite{Takayanagi:2011zk} for the gravity dual of BCFT (the proposal passes many checks). Throughout this section we consider two complementary regions, call them $A$ and $B$, in the thermal state of a holographic CFT on an interval. We assume for simplicity that the size of region $A$ is always less than or equal to the size of region $B$ and that $A$ and $B$ together give the whole CFT.

The basic proposal for the gravity dual of BCFT is to solve Einstein's equations in the presence of an ``end of the world brane" which terminates the bulk spacetime and which ends on the boundary of the boundary, i.e. the boundary of the CFT spacetime. In the simplest case, this brane is described just by a tension $\mathcal{T}$. One then solves the bulk Einstein equations plus the equation of motion of the brane to find a bulk spacetime with an asymptotic boundary and a bulk termination at the brane. The rules for calculating entanglement entropy are the same, but with the extra proviso that the end of the world brane never contributes.

Practically speaking, for the simple case of three dimensional Einstein gravity which we consider here, the geometry is either described by a part of empty AdS or a part of the BTZ black hole. At low or zero temperature, the dominant saddle point is the AdS geometry. The metric of AdS may be taken to be
\begin{equation}
ds^2 = \lads^2\left(-\frac{dt^2}{z^2} + \frac{dz^2}{z^2 h(z)} + \frac{h(z) dx^2}{z^2} \right)
\end{equation}
where $h(z) = 1 - z^2/z_0^2$ and $x$ is periodic with period $2\pi z_0$. The terminating brane is denoted $Q$ and is described by the curve \cite{Takayanagi:2011zk}:
\begin{equation}
    Q: x(z) - x(0) = z_0 \tan^{-1} \frac{\lads \mathcal{T} z}{z_0 \sqrt{h(z) - \lads^2 \mathcal{T}^2}}.
\end{equation}
The turning point of this curve is at $z=z_0 \sqrt{1 - \lads^2 \mathcal{T}^2}$ and its mirror continues after the turning point. The total length of the boundary interval is thus
\begin{equation}
    2 z_0 \tan^{-1} \infty = \pi z_0.
\end{equation}

As the temperature is increased, the system experiences a first order Hawking-Page transition from an AdS geometry to a BTZ black hole geometry. The black hole geometry may be written as
\begin{equation}\label{BTZSchwarz}
    ds^2 = \lads^2\left(-\frac{f(z) dt^2}{z^2} + \frac{dz^2}{z^2 f(z)} + \frac{dx^2}{z^2} \right)
\end{equation}
where $f(z) = 1 - z^2/z_H^2$ and the temperature is $\beta = T^{-1} = 2\pi z_H$. The terminating brane is now
\begin{equation}
    Q: x(z) - x(0) = z_H \sinh^{-1} \frac{\lads \mathcal{T} z}{z_H \sqrt{1-\lads^2 \mathcal{T}^2}}.
\end{equation}
The length of the boundary at $z=0$ is still written as $\pi z_0$, and for positive tension $\mathcal{T}$ the horizon $z=z_H$ includes more of the $x$ coordinate. By analyzing the free energy of the AdS and BTZ saddle points, one can show that the Hawking-Page transition occurs when
\begin{equation}
    \frac{\pi z_0}{\beta} = \sqrt{\frac{1}{4} + \left(\frac{1}{\pi}\tanh^{-1} \lads \mathcal{T}\right)^2} - \frac{1}{\pi}\tanh^{-1} \lads \mathcal{T}.
\end{equation}
For example, if the string tension goes to zero, then the phase transition occurs when $z_0=z_H$. By contrast, as the string tension gets large, the phase transition occurs at larger and larger $\beta$.

Now to study the entanglement of purification of as a function of the relative size of $A$ and $B$ we must consider two variables. Fixing the total size, we must first determine, as a function of temperature, whether we are in the AdS or BTZ phase. Then, given the geometry, we must perform the minimization over curves according to the rules discussed above to find the holographic entanglement of purification. This procedure is somewhat involved, so we will not consider the general case here (we anyway do not expect an extremely detailed correspondence between the spin chain and holographic model - for example, the spin chain has no phase transition while the holographic model does). We will consider a few limits and special cases.

First, consider the limit of high temperature (or large interval size) and the case where $A$ is just less than half the total system size, $|A| = \pi z_0/2$. In this limit the boundary effects are mostly irrelevant, at least at finite temperature, and the calculations are simplified. The dominant geometry is the BTZ black hole and the minimal cross-section of the $AB$ entanglement wedge is simply given by a curve which drops vertically from $z=\epsilon$ (the regulated asymptotic boundary) to $z=z_H$. The length of this curve in Planck units is the holographic entanglement of purification; we find
\begin{equation}
    E_p  = \frac{\lads}{4 G_N} \log \frac{\beta}{\pi \epsilon} = \frac{c}{6}  \log \frac{\beta}{\pi \epsilon} .
\end{equation}
To remove the dependence on the cutoff, it is natural to compare to the ground state entropy of $A$. On general CFT grounds, the ground state entropy is given by
\begin{equation}
    S_{gs} = \frac{c}{6} \log \left(\frac{2 L}{\pi \epsilon} \sin \frac{\pi L_{A}}{L}\right) + \log g
\end{equation}
where $\log g$ is the boundary entropy and $L=\pi z_0$ is the total length. In holographic BCFT, the boundary entropy is related to the string tension via
\begin{equation}
    \log g =\frac{c}{6} \tanh^{-1} \lads\mathcal{T}.
\end{equation}
When $L_{A}=L/2$, the ground state entropy is $S_{gs} = \frac{c}{6} \log \frac{2 L}{\pi \epsilon}+\log g$. Hence the UV finite scaling form reads
\begin{equation}
    e^{E_p - S_{gs}} = \frac{1}{g} \left(\frac{\beta}{2L}\right)^{c/6}.
    \label{eq:bcftscaling}
\end{equation}

Another interesting comparison is to the entanglement between $AA'$ and $BB'$ (where $A'$ and $B'$ are the mirrors of $A$ and $B$ in the purifier) in the thermofield double state. This entanglement is actually just twice $E_p$ in this limit. Since the required bond dimension is $\chi \sim e^{E_p}$, the minimal purification is predicted to require approximately the square root of the bond dimension needed for the thermofield double state. Note that in this limit, the holographic entanglement of purification is also approximately the mutual information, so if the holographic prescription is correct, then the lower bound on $E_p$ is close to being reached.

It is also possible to study $E_p$ as a function of the size of $A$. If the system is in the thermal AdS phase, then $E_p=S(A)$ provided $A$ is less than half the total system. In the holographic model, what is in essence happening is that the dual gauge theory is confined and the system is essentially in its ground state except for a few thermal modes. Hence the large $N$ part of the entanglement is like that of a pure state. If the system is in the BTZ phase, then $E_p = S(A)$ again for sufficiently small $A$, but beyond a critical size of $A$, $E_p$ saturates to the value
\begin{equation}
    E_p = \frac{c}{6} \log \frac{\beta}{\pi \epsilon}
\end{equation}
as discussed above. These two features, tracking the entropy of $A$ for small $A$ and rapidly saturating for large $A$, are strikingly similar to the spin chain data, at least for sufficiently high temperature.

We conclude this discussion by working out the simplest example in slightly more detail. We consider the case of vanishing string tension, $\mathcal{T}\rightarrow 0$. Note that in this limit the boundary entropy goes to zero,
\begin{equation}
    \lim_{\mathcal{T}\rightarrow 0} \log g = \lim_{\mathcal{T}\rightarrow 0} \frac{c}{6} \tanh^{-1} \lads \mathcal{T} = 0.
\end{equation}
Similarly, the Hawking-Page transition occurs for $z_0 = z_H$. Geometrically, the key simplifying feature is that the $Q$ boundary is now essentially vertical, i.e. independent of $z$. We already argued on general grounds that at low temperatures the holographic entanglement of purification is simply $E_p = S(A)$. Therefor let us consider the high temperature case.

In the high temperature phase, the entanglement entropy of $A$ for any region $A$ less than half the system size can be obtained by using a doubling trick. The entropy of a segment terminating at the boundary is simply one half the entropy of a segment of twice the size without the boundary. This is correct in the limit where $Q$ is vertical. Thus if $A$ is an interval of length $L_A$ then
\begin{equation}
    S(A) = \frac{1}{2} S_{\text{no boundary}}(2 L_A) =  \frac{c}{6} \log \left( \frac{\beta}{\pi \epsilon} \sinh \frac{2 \pi L_A}{\beta} \right).
\end{equation}
The entanglement of purification is given by the minimum length among two candidate curves, the minimal curve for $A$ and the vertical segment running from $z=\epsilon$ to $z=z_H$. For large $L_A$, the vertical segment dominates. For small $L_A$, the minimal curve for $A$ dominates. By equating the entropy of $A$ with the length of the vertical segment, we see that the two curves exchange dominance when
\begin{equation}
    \sinh \frac{2 \pi L_A}{\beta} = 1
\end{equation}
or
\begin{equation}
    \frac{L_A}{\beta}=\frac{\log( 1+\sqrt{2})}{2\pi}\approx .140...
\end{equation}
Thus we have
\begin{align}
    E_p & = \frac{c}{6} \log \left( \frac{\beta}{\pi \epsilon} \sinh \frac{2 \pi L_A}{\beta} \right) \,\,\,\, (L_A/\beta < .140...) \nonumber \\
    & = \frac{c}{6} \log \frac{\beta}{\pi \epsilon} \,\,\,\, (L_A/\beta > .140...).
\end{align}

If $L_A$ is half the total system size, $L_A=\pi z_0/2$, then the switch occurs at
\begin{equation}
    \frac{z_0}{z_H} = \frac{2\log( 1+\sqrt{2})}{\pi} \approx .561...
\end{equation}
However, the Hawking-Page transition occurs at $z_0/z_H=1$, so the geometry switches to AdS before the change of minimal curve can occur in the BTZ geometry. Hence the scaling function $e^{E_p-S_{gs}}$ has the following form in the tensionless limit,
\begin{equation}
e^{E_p - S_{gs}} = \begin{cases}
      \left(\frac{\beta}{2L}\right)^{c/6} & \frac{\beta}{L} < 2 \\
      1 & 2 < \frac{\beta}{L}
   \end{cases}
\end{equation}
By accident, in this limit the scaling function is actually continuous across the Hawking-Page transition.\\
We also consider the case $L_{A} = L/4$ and zero brane tension. In this case the Hawking-Page transition still occurs at $\beta/L = 2$. But at high temperature ($\beta/L < 2$), we have a competition between the surface that drops vertically into the bulk and the one that terminates on $Q$, and they exchange dominance around $\frac{\beta}{L} \approx 1.782...$. The scaling form is found to be:
\begin{equation}
e^{E_p - S_{gs}} = \begin{cases}
      \left(\frac{\beta}{\sqrt{2} L}\right)^{c/6} & \frac{\beta}{L} < 1.782... \\
      \left[ \frac{\beta}{\sqrt{2}L} \sinh{\left( \frac{\pi\beta}{2L} \right)} \right]^{c/6} & 1.782... < \frac{\beta}{L} < 2 \\
      1 & 2< \frac{\beta}{L}
   \end{cases}
\end{equation}

\section{Random stabilizer tensor networks}\label{Sec:TensorNetwork}

Having motivated the holographic prescription in part using tensor networks, in this section we discuss one concrete tensor network computation of $E_p$. Unlike the previous two models, here our results are rigorously correct. Based on the relationship between tensor networks and the AdS/CFT correspondence, there has been considerable interest in designing tensor networks which obey the network version of the RT formula. Random stabilizer tensor networks are one class that obeys the RT formula. Here we show, using the results of Ref.~\cite{Nezami:2016zni}, that the entanglement of purification can be easily calculated in random stabilizer tensor networks and that it reduces to approximately $\frac{1}{2} I(A:B)$.

Consider a connected graph $(V,E)$ and choose a subset $V_\partial$ of the vertices called ``boundary vertices". These vertices are the analog of the CFT degrees of freedom which live on the boundary in the AdS/CFT correspondence. The remaining vertices are called ``bulk vertices" and they are the analog of the gravity degrees of freedom in the AdS/CFT correspondence. We associate a tensor $|V_x\rangle$ to each vertex $x \in V_b$ and a maximally entangled state $|e \rangle$ to each edge $e \in E$. The bond dimension is taken to be $\chi$ for all bonds so that $|e\rangle = \frac{1}{\sqrt{\chi}} \sum_{i=0}^{\chi-1}\ket{ii}$ and $|V_x \rangle$ is a tensor on a $\chi^{\text{deg}(x)}$ dimensional space where $\text{deg}(x)$ is the degree of vertex $x$. The final pure quantum state on $V_\partial$ is
\begin{equation}
    |\psi_\partial \rangle = \left( \bigotimes_{x \in V_b} \langle V_x | \right) \bigotimes_{e \in E} |e \rangle.
\end{equation}

The above construction is quite general. A stabilizer state can be constructed by first taking the bond dimension to be $\chi = p^N$ for prime $p$. Then the maximally entangled states are stabilizer states. If the vertex tensors are also taken to be stabilizer states, then the resulting pure state on $V_\partial$ is also a stabilizer state. A random stabilizer state is obtained by drawing the tensors $|V_x \rangle$ uniformly at random from the set of all stabilizer states of the relevant dimension.

One of the main results of Ref.~\cite{Hayden:2016cfa} is that such random stabilizer states obey the network RT formula. Given a subset $A$ of $V_\partial$, the entropy of $A$ in state $|\psi_\partial \rangle$ is given by the minimal number of bonds in the network which must be cut to isolate $A$,
\begin{equation}
    S(A) \approx N \log p \times |\text{minimal cut}|.
\end{equation}
For the remainder of this section, all entropies will be measured in units of $\log p$, so the RT formula reads $S(A) = N |\text{minimal cut}|$. This result fully characterizes the bipartite entanglement in random stabilizer tensor networks.

Recently, progress has also been made on properties of multipartite entanglement in random stabilizer states. Consider a tripartite stabilizer state $|\psi \rangle_{ABC}$. It is known that, up to local unitary transformations, the entanglement content of such a state is given by Bell pairs and GHZ states \cite{PhysRevA.84.052306,doi:10.1063/1.2203431}. Denote the Bell pair by
\begin{equation}
|\Phi \rangle_{AB}= \frac{1}{\sqrt{p}} \sum_{i=0}^{p-1} |i\rangle_A | i \rangle_B,
\end{equation}
and the GHZ state by
\begin{equation}
    |\text{GHZ}\rangle_{ABC} = \frac{1}{\sqrt{p}} \sum_{i=0}^{p-1} |i \rangle_A | i \rangle_B |i \rangle_C.
\end{equation}
Note that these states do not depend on $N$, i.e. they represent elementary units of entanglement. In this notation, the statement is that for any tripartite pure state there exist local unitaries $U_A$, $U_B$, and $U_C$ and factors $A_i$, $B_i$, and $C_i$ of the $A$, $B$, and $C$ Hilbert spaces such that
\begin{equation}
    U_A U_B U_C |\psi\rangle_{ABC} = \left(|\Phi\rangle_{A_1 B_1}\right)^c \left(|\Phi \rangle_{B_2C_1}\right)^a \left(|\Phi \rangle_{A_2 C_2}\right)^b (\left |\text{GHZ}\rangle_{A_3 B_3 C_3} \right)^g
\end{equation}
up to unentangled states.

Given this form, it is easy to calculate the entropy of any region, say $A$:
\begin{equation}
    S(A) = b+c+g.
\end{equation}
Similarly, the mutual information is
\begin{equation}
    \frac{1}{2} I(A:B) = c + \frac{g}{2}.
\end{equation}
Finally, using results outlined in the introduction plus the fact that the state of $AB$ reduces to products of decoupled mixed states, Bell pairs, and purely classically correlated states (arising from GHZ), the entanglement of purification can be calculated:
\begin{equation}
    E_p(A:B) = c+ g.
\end{equation}

Now, in the limit of large $N$, the numbers $a$, $b$, and $c$ scale with $N$ while the number $g$ is order one \cite{Nezami:2016zni}. Hence it follows that
\begin{equation}
    E_p(A:B) = \frac{1}{2} I(A:B) + \frac{g}{2} \approx \frac{1}{2} I(A:B).
\end{equation}
In other words, in random stabilizer tensor networks, the entanglement of purification is approximately the lower bound of one half the mutual information. This is in contrast to the holographic proposal, where $E_p$ and $\frac{1}{2}I$ could differ by a large amount. Indeed, we could have considered an analog of the holographic proposal for random stabilizer tensor networks, but this proposal would be wrong in general.

The random stabilizer tensor network result does highlight an important caveat in the holographic discussion. Since such networks obey the RT formula, any property derived from RT is also obeyed in such networks. Similarly, one can show that in holographic systems which obey the RT formula, the lower bound of $\frac{1}{2} I(A:B)$ is also consistent with all properties of $E_p$. Hence it is prudent to emphasize that it is possible the holographic answer is simply one half the mutual information; however, it must be similarly emphasized that the entanglement structure of holographic states is known to be more complex than that of stabilizer states, e.g. the spectrum of density matrices is not flat.

One final note is appropriate. There are other classes of tensor network states that obey the network version of the RT formula, e.g. some tensor networks made of perfect tensors and random tensor network states. Especially in the case of random tensor networks, it is natural to conjecture that the holographic prescription giving $E_p$ in terms of the entanglement wedge cross section generalizes to its network version. It would be very interesting to prove or refute this conjecture in the class of random tensor networks.

\section{Holographic proposal: general formulation and properties} \label{Sec:HolographyGeneral}

In this section we return to our holographic proposal and discuss some general features of it. First, we generalize it to time-dependent situations. Then we discuss some interesting features of the proposal, especially the case when $E_{ph}$ undergoes a first order phase transition. Finally, we show that our proposal in the time-independent case obeys all the properties of $E_p$ listed in the technical introduction. The time dependent case is more complex, and depends in principle on the actual dynamics of the theory, so we leave it for future work.

\subsection{Holographic proposal: time-dependent case}
\begin{figure}[H]
    \centering
    \includegraphics[width=4.5cm]{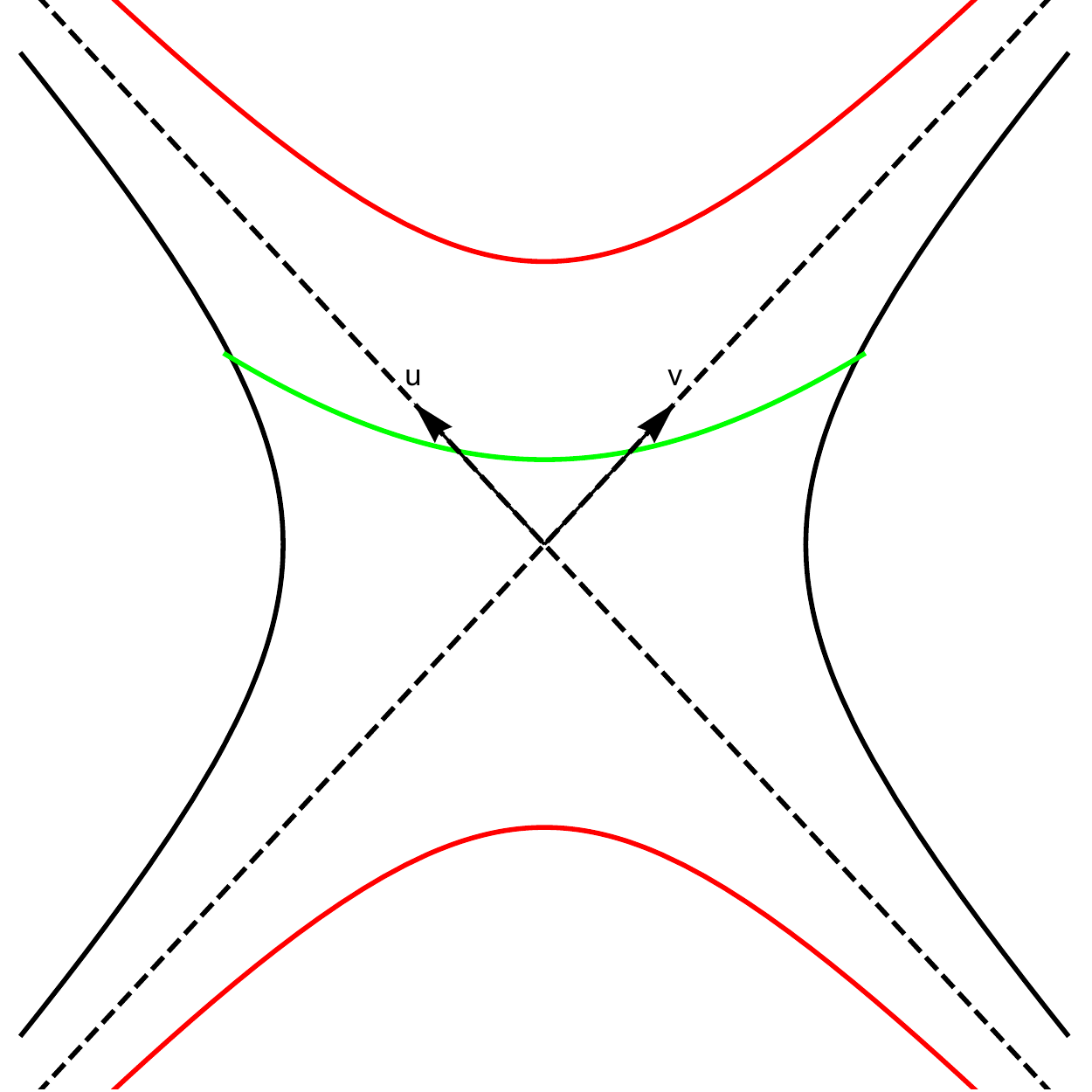}
     \includegraphics[width=4.5cm]{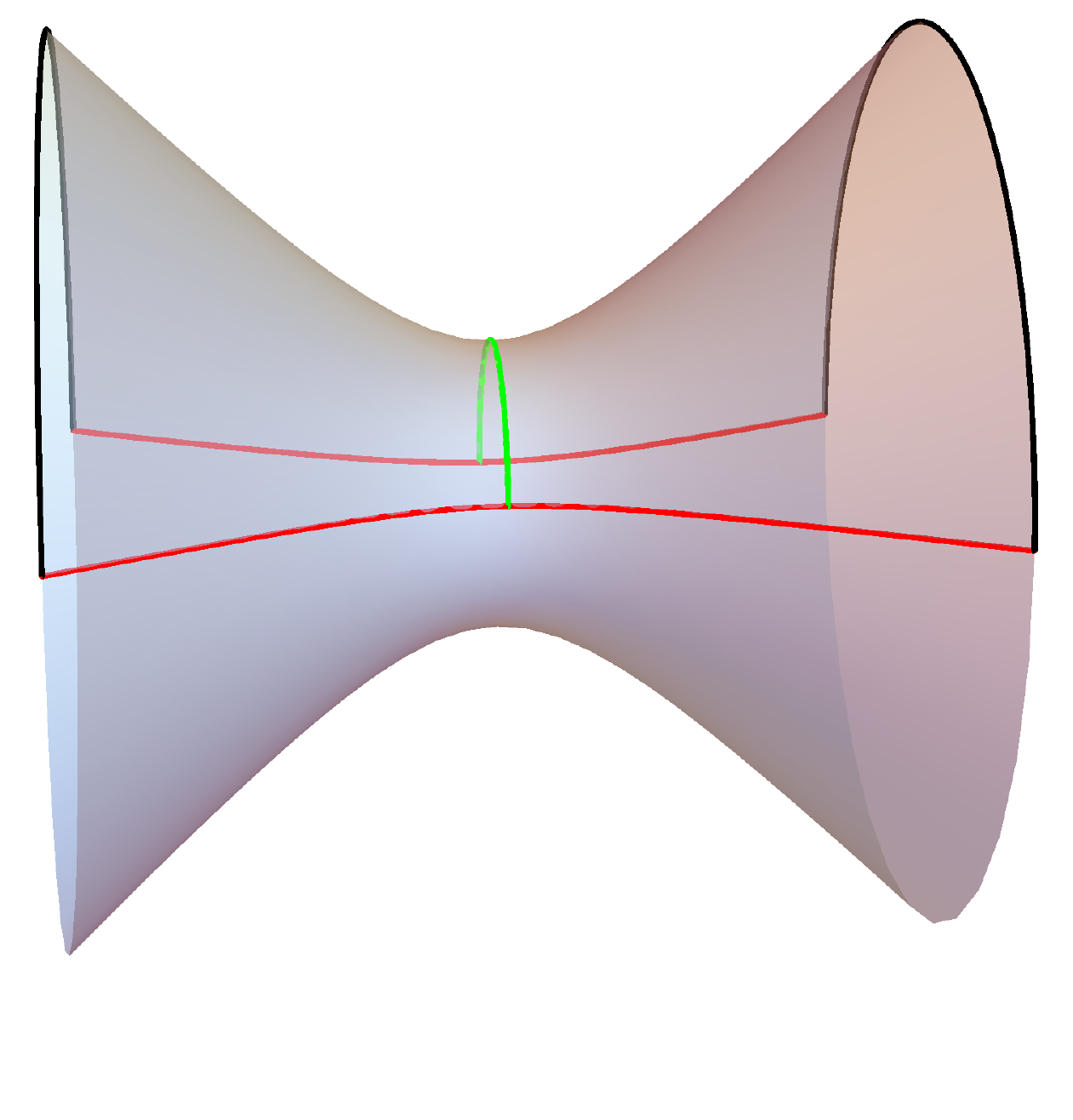}
    \caption{Left: The HRT surface (green) is a pair of geodesics crossing the wormhole anchored at the same boundary time on the left and on the right. Right: the topology of a spatial slice is that of a cylinder. We draw schematically a spatial slice which contains the $E_{p}$ surface.}
    \label{fig:2sidedBTZ}
\end{figure}

\begin{figure}[H]
    \centering
    \includegraphics[width=9cm]{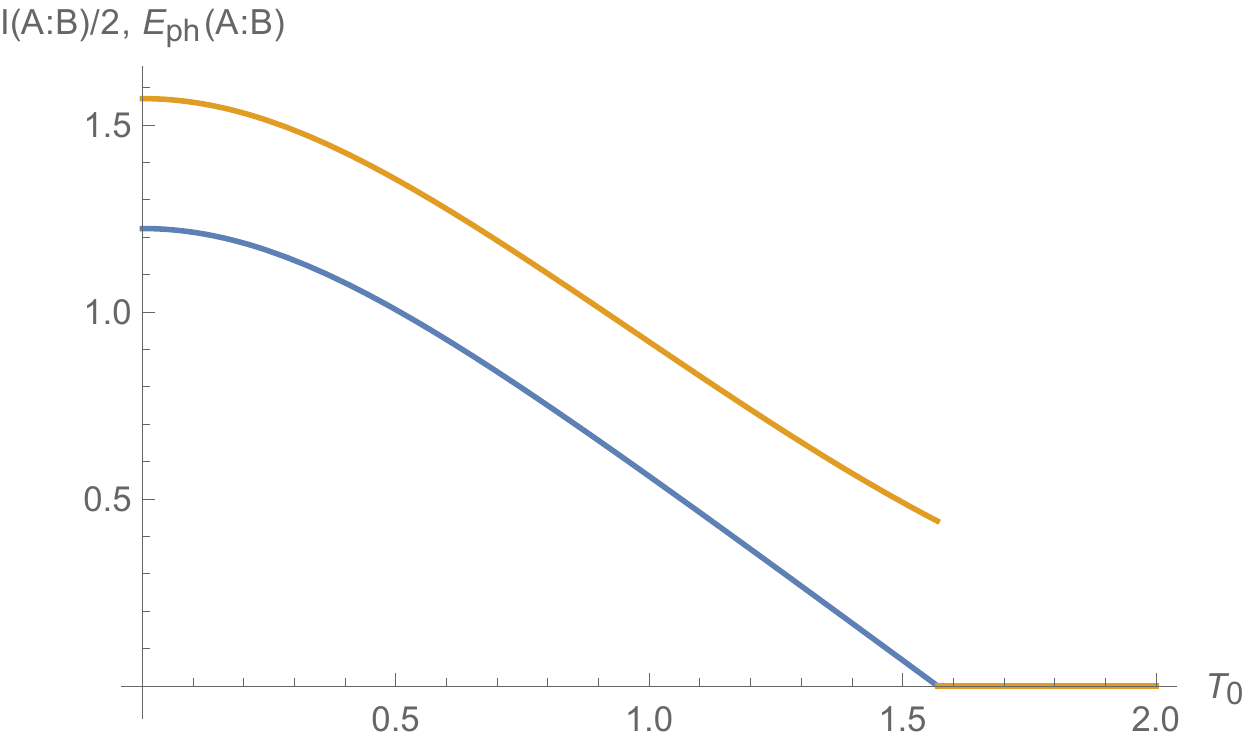}
    \caption{We plot the $E_{ph}$ (in orange) and half the mutual information (in blue) as a function of $T_{0}$, with $z_{H} = 1/2$, $G_{N}=1$ and $\lads = 1$. The mutual information becomes zero at around $T_{0} = 1.568$.}
    \label{fig:EphTimeDependent}
\end{figure}

Our proposal for the holographic entanglement of purification can be generalized to a time-dependent setting in a straightforward manner. Given two boundary regions $A$ and $B$, the $E_{ph}{(A:B)}$ is the length of the shortest of all extremal surfaces in the entanglement wedge that separates $A$ from $B$, and this extremal surface is allowed to terminate on the HRT surface \cite{Hubeny:2007xt} which we will call $\Gamma$. Put differently, we think of the entanglement wedge as a new spacetime with spatial boundary $A \cup B \cup \Gamma$. Then we again consider all partitions of $\Gamma$ into $A'$ and $B'$ and minimize the entropy of $AA'$, as computed by HRT, over the choice of $A'$. This proposal for time-dependent $E_{ph}$, of course, reduces to the bottleneck of the entanglement wedge in the static case.\\\\
For example, consider the case of the 2-sided BTZ black hole. The boundary consists of 2 circles, and we want to compute $E_{ph}{(A:B)}$ where $A$ and $B$ are each half of each boundary circle from $\phi=0$ to $\phi=\pi$, at the same boundary time \footnote{Note that the Schwarzschild time increases downward on the left boundary and upward on the right boundary. When we say ``same boundary time'', we mean the boundary time on the left is the negative of the boundary time on the right.}. A similar setup was considered in \cite{Hartman:2013qma} to study the time dependence of the entanglement entropy. First, we find the HRT surface, which we will denote $\Gamma$: $\Gamma$ a pair of spacelike geodesic crossing the wormhole connecting $A$ to $B$. The HRT surface is disconnected and consists of 2 connected component, as depicted in Figure \ref{fig:2sidedBTZ}. By symmetry, the $E_{ph}$ should be the geodesic distance between the two midpoints of the connected components of $\Gamma$. We schematically depict this in Figure \ref{fig:2sidedBTZ}.
Using the fact that BTZ is a quotient of $AdS_{3}$, one can work out an analytical formula for the $E_{ph}$ as a function of the boundary time $T_{0}$ (by boundary time, we mean the Schwarzschild or Killing time on the boundary). In Kruskal coordinates, the BTZ metric reads:
\begin{equation}
    ds^{2} = \frac{-4\lads^{2}dudv + R^{2}(1-uv)^{2}d\phi^{2}}{(1+uv)^{2}}
\end{equation}
with $\phi \sim \phi + 2\pi$. We need the geodesic distance between any two spacelike-separated points $X_{1}=(u_{1},v_{1},\phi_{1})$ and $X_{2}=(u_{2},v_{2},\phi_{2})$ in the BTZ spacetime \cite{Morrison:2012iz}:
\begin{equation}\label{DisBTZBulk}
    D{(X_{1},X_{2})} = \lads \mathrm{arccosh}{[-\Theta{(X_{1},X_{2})}]}
\end{equation}
with
\begin{equation}
    \Theta{(X_{1},X_{2})} = -\frac{2(u_{1}v_{2}+v_{1}u_{2})+(1-u_{1}v_{1})(1-u_{2}v_{2})\cosh{\left(\frac{\lads (\phi_{1}-\phi_{2})}{z_{H}}\right)}}{(1+u_{1}v_{1})(1+u_{2}v_{2})}
\end{equation}
In particular, for two points on the boundary $X_{1}=(t_{1},\phi_{1})$ and $X_{2} = (t_{2},\phi_{2})$ (in the Schwarzschild coordinates of equation (\ref{BTZSchwarz}) with the renaming of the coordinate $x \rightarrow \phi$), we have the distance formula:
\begin{equation}\label{DisBTZBoundary}
    D{(X_{1},X_{2})} = \lads \ln{\left[ \frac{-2\Theta{(X_{1},X_{2})}}{\epsilon^{2}} \right]}
\end{equation}
with
\begin{equation}
    \Theta{(X_{1},X_{2})} = z_{H}^{2} \left[\pm \cosh{\left(\frac{t_{1}-t_{2}}{z_{H}} \right)} - \cosh{\left( \frac{\lads (\phi_{1}-\phi_{2})}{z_{H}} \right)} \right]
\end{equation}
where the sign of $\pm$ is plus if the two points belong to the same boundary, and minus if they belong to different boundaries, and $\epsilon$ is a regulator defined by integrating the geodesic up to the near-boundary hyperbola $uv = -1 + 2\epsilon/z_{H}$. We now consider the 4 points $a$,$b$,$c$, and $d$ which are the endpoints of $A$ and $B$ (the black semicircles on the right panel of Figure \ref{fig:2sidedBTZ}). Their coordinates are:
\begin{equation}
    a = (t=-T_{0},\phi=0)
\end{equation}
\begin{equation}
    b = (t=-T_{0},\phi=\pi)
\end{equation}
\begin{equation}
    c = (t=T_{0},\phi=0)
\end{equation}
\begin{equation}
    d = (t=T_{0},\phi=\pi)
\end{equation}
Here $a$, $b$ lie on the left boundary and $c$, $d$ lie on the right boundary, and the time coordinates of $a$ and $b$ are negative because the time coordinate increases downward on the left boundary. Using the distance formula (\ref{DisBTZBoundary}) above, we can find $S(A) = S(B) = \frac{D(a,b)}{4G_{N}} = \frac{D(c,d)}{4G_{N}}$ and $S(AB) = 2 \frac{D(a,c)}{4G_{N}} = 2 \frac{D(b,d)}{4G_{N}}$:
\begin{equation}
    S{(A)} = S{(B)} = \frac{\lads}{2G_{N}} \ln{\left[ \frac{2z_{H}}{\epsilon} \sinh{\left( \frac{\pi \lads}{2z_{H}} \right)} \right]}
\end{equation}
\begin{equation}
    S{(AB)} = \frac{\lads}{G_{N}} \ln{\left[ \frac{2z_{H}}{\epsilon} \cosh{\left( \frac{T_{0}}{z_{H}} \right)} \right]}
\end{equation}
Note that $S(A)$ and $S(B)$ are independent of $T_{0}$. This is because both these RT surfaces lie on a spatial slice of fixed Schwarzschild time (which goes through the bifurcation surface of the black hole), and the metric is static in this time coordinate. The mutual information is nonzero from time $T_{0}=0$ to:
\begin{equation}
    T_{*} = z_{H} \mathrm{arccosh}{\left[ \sinh{\left( \frac{\pi \lads}{2z_{H}} \right)} \right]}
\end{equation}
at which point there is a phase transition and the mutual information jumps to zero. During the time $0 \leq T_{0} \leq T_{*}$, the mutual information is given by:
\begin{equation}
    I{(A:B)} = \frac{\lads}{G_{N}} \ln{\left[ \sinh{\left( \frac{\pi \lads}{2z_{H}} \right)} \mathrm{sech}{\left( \frac{T_{0}}{z_{H}} \right)} \right]}
\end{equation}
As for the $E_{ph}$, it is given by the geodesic distance between the midpoint of the component of $\Gamma$ connecting $a$ to $c$, and the midpoint of the component connecting $b$ to $d$. These two midpoints are located at $(u,v)$ coordinates given by:
\begin{equation}
    u = v = \tanh{\left( \frac{T_{0}}{2z_{H}} \right)}
\end{equation}
At $T_{0}=0$, the midpoint of the HRT surface is the bifurcation circle of the black hole ($u=v=0$). As $T_{0} \rightarrow \infty$, the midpoint approaches the singularity $(u=v=1)$. Using the distance formula (\ref{DisBTZBulk}), we find for the $E_{ph}$:
\begin{equation}
E_{ph}{(T_{0})} = \frac{\lads}{4G_{N}} \mathrm{arccosh}{\left\{ 1 + \left[ \cosh{ \left( \frac{\pi \lads}{z_{H}} \right)} - 1\right] \mathrm{sech}^{2}{\left( \frac{T_{0}}{z_{H}} \right)} \right\} }
\end{equation}
In particular, at boundary time $T_{0}=0$ the $E_{ph}$ is equal to half the circumference of the bifurcation circle of the black hole (divided by $4G$). We plot in Figure (\ref{fig:EphTimeDependent}) the time evolution of the $E_{ph}$ and (half) the mutual information. Note that, as expected, the $E_{ph}$ is greater than or equal to half the mutual information.\\
A peculiar feature of the $E_{ph}$ in this case, as can be seen from Figure \ref{fig:EphTimeDependent}, is that even as the mutual information approaches zero continuously at the phase transition, the $E_{ph}$ remains finite and then jumps discontinuously to zero (with the difference between $E_{ph}$ and half the mutual information approximately constant in time until the phase transition). This behavior is somewhat counterintuitive, as one would expect the mutual information and the entanglement of purification to behave similarly to each other. Nevertheless this is also what occurs for 2 non-adjoint boundary intervals in empty $AdS$: when the entanglement wedge transitions from being connected to being disconnected, the mutual information approaches zero continuously while the $E_{ph}$ jumps discontinuously to zero. It would be interesting to understand this phenomenon in more details. In particular, it would be nice to construct explicit quantum states which have close to zero mutual information but nonzero $E_{p}$.\\

\subsection{Holographic check of inequalities}\label{Sec:HolographicIneq}
In this section, we show that $E_{ph}$ satisfies the inequalities mentioned in the technical introduction. We will first go through each inequality and check its validity in time-independent backgrounds. Then we will generalize the arguments to the time-dependent case at the end.\\
\paragraph{Upper bound by entanglement entropy.} First we check the upper bound (\ref{EPUpperBound}). For 2 adjacent intervals in AdS, this bound is trivially satisfied because the $E_{ph}$ is UV-divergent at one endpoint but each RT surface for $S(A)$ and $S(B)$ diverges at both endpoints. For 2 non-adjacent intervals, the bound is also trivially true since the entanglement entropy diverges but the $E_{ph}$ is finite.\\
The BTZ case is more subtle. Consider for example the symmetrical case where $A$ and $B$ are each half the boundary on one side (their half-widths are both $\pi/2$). The $E_{ph}$ has already been computed:
\begin{equation}
    E_{ph}{(A:B)} = \frac{\lads}{2G_{N}} \log{\left( \frac{2r_{c}}{r_{+}} \right)}
\end{equation}
and the entanglement entropies are:
\begin{equation}
    S{(A)} = S{(B)} = \frac{\lads}{2G_{N}} \log{\left( \frac{2r_{c}}{r_{+}} \sinh{\left( \frac{\pi r_{+}}{2\lads} \right)} \right)}
\end{equation}
The question of whether $E_{p}{(A:B)} \leq S{(A)}$ then depends on the sign of the quantity $ 2\lads \log{\left(\sinh{\frac{\pi r_{+}}{2\lads}}\right)}$. This quantity could be of either sign, depending on the size of the horizon relative to $L$, but we can invoke a thermodynamic argument to eliminate the negative case. Recall that the $BTZ$ black hole undergoes the Hawking-Page transition to thermal AdS when the horizon is smaller than the $AdS$ lengthscale, and only large black holes (with $r_{+} > L$) are thermodynamically stable. For large black holes, we have that $2\lads \log{\left(\sinh{\frac{\pi r_{+}}{2\lads}}\right)} > 0$ and the upper bound by the entanglement entropy is satisfied.

\paragraph{Monotonicity.} The monotonicity property is quite intuitively clear. For 2 adjacent intervals in $AdS_{3}$, recall formula (\ref{EphAdjacent}) for the $E_{ph}$. If we differentiate this formula with respect to $\alpha_{1}$, we have:
\begin{equation}
    \frac{\partial E_{ph} (\alpha_{1},\alpha_{2})}{\partial \alpha_{1}} = \frac{\lads}{4G_{N}} \csc{(\alpha_{1})}\sin{\alpha_{2}}\csc{(\alpha_{1}+\alpha_{2})}
\end{equation}
Since both $\alpha_{1}$ and $\alpha_{2}$ are in the range $(0,\pi/2)$, the quantity above is always positive. This means the $E_{ph}$ indeed increases monotonically with $\alpha_{1}$ at fixed $\alpha_{2}$. Similarly for $\alpha_{2}$. For non-adjacent intervals in $AdS_{3}$ as well as adjacent or non-adjacent intervals in BTZ, one can similarly differentiate the $E_{ph}$ formulae and check that it is positive.

\begin{figure}[H]
    \centering
     \includegraphics[width=5cm]{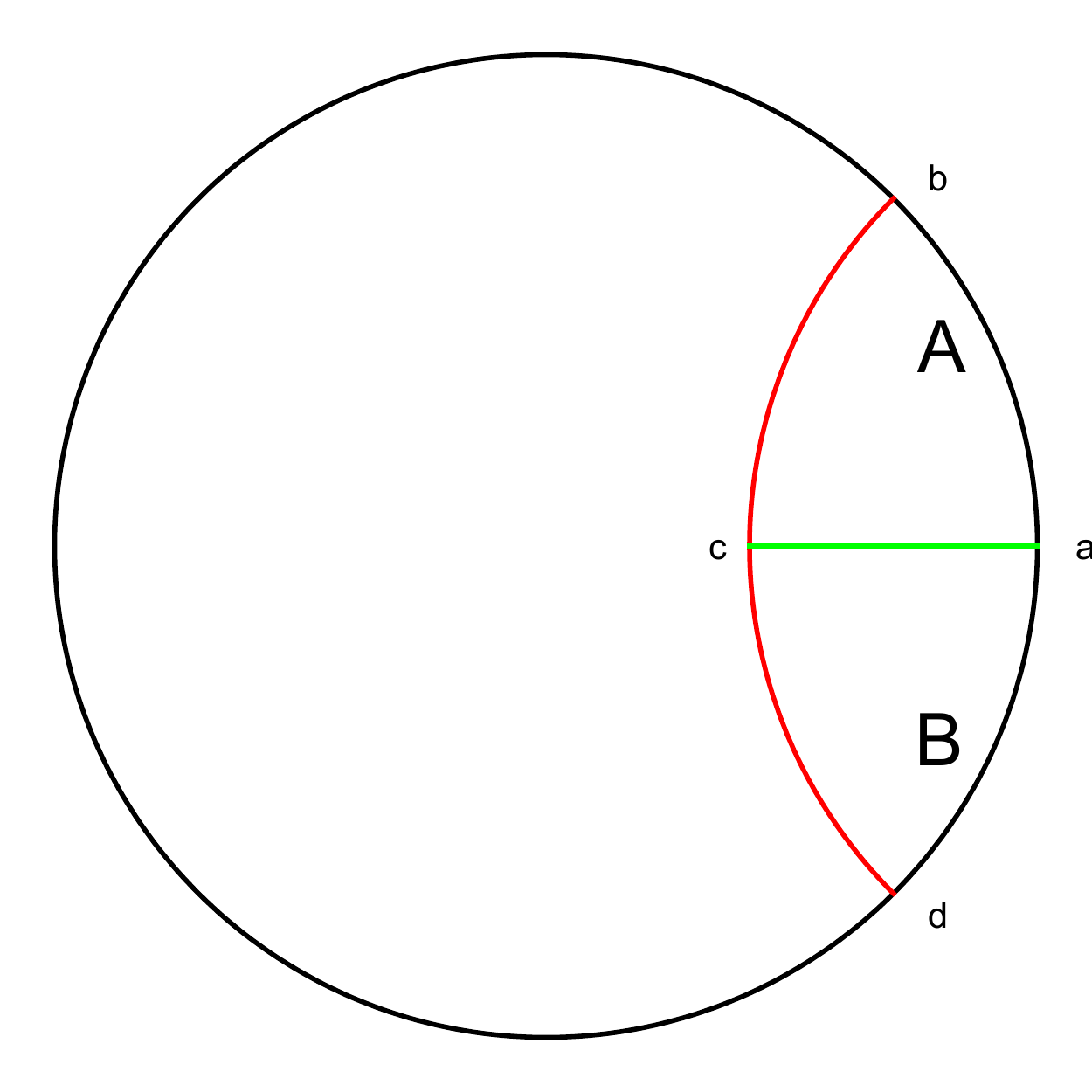}
    \caption{Graphical proof of the lower bound $E_{ph}{(A:B)} \geq \frac{1}{2}I{(A:B)}$ for two adjacent intervals.}
    \label{fig:LowerBoundPf}
\end{figure}

\paragraph{Lower bound by the mutual information.} Next, we check the bound (\ref{EPLowerBound}). For 2 adjacent intervals, the lower bound is a simple consequence of Riemannian geometry, as illustrated in Figure (\ref{fig:LowerBoundPf}). Let $a$, $b$, $c$ and $d$ be points as labelled on the figure. We will denote by $(ab)$ the length of the geodesic connecting $a$ and $b$ etc.
We then have:
\begin{equation}
    E_{ph}{(A:B)} = \frac{(ac)}{4G_{N}}
\end{equation}
\begin{equation}
    I{(A:B)} = \frac{1}{4G_{N}} \left[(ab) + (ad) - (bc) - (cd) \right]
\end{equation}
But, by definition of a geodesic, we also have $(ab) < (ac) + (bc)$ and $(ad) < (ac) + (cd)$. Plugging the two inequalities above into $I{(A:B)}$ above, we find
\begin{equation}
I{(a:b)} < \frac{(ac)}{2G_{N}} = 2 E_{ph}{(A:B)}
\end{equation}
thus proving the bound. Similar proofs can be constructed for two non-adjacent intervals as well as the BTZ black hole in a straightforward way, as well as for other asymptotically AdS geometries.\\
Even though we have established the lower bound, it is still interesting to explicitly compute the difference between $E_{ph}$ and half the mutual information in a few simple cases. For two arbitrary adjacent intervals of half-widths $\alpha_{1}$ and $\alpha_{2}$, the $E_{ph}$ and mutual information are:
\begin{equation}
    E_{ph}{(A:B)} = \frac{\lads}{4G_{N}} \log{(2 \csc{(\alpha_{1}+\alpha_{2})} \sin{\alpha_{1}} \sin{\alpha_{2}})} + \frac{\lads}{4G_{N}} \log{\left( \frac{2R_{c}}{\lads} \right)}
\end{equation}
\begin{equation}
    \frac{1}{2}I{(A:B)} = \frac{\lads}{4G_{N}} \log{\left( 2 \csc{(\alpha_{1}+\alpha_{2})} \sin{\alpha_{1}} \sin{\alpha_{2}} \right)} + \frac{\lads}{4G_{N}} \log{\left( \frac{R_{c}}{\lads} \right)}
\end{equation}
Comparing the two expressions above, we find that this latter is larger than half the mutual information by an amount $\lads \log{2}$.\\
Next, consider 2 non-adjacent intervals. For the simple special case where $A$ and $B$ have the same size $\alpha$ and are diametrically opposite each other (with $\alpha$ sufficiently large so that the entanglement wedge is connected), the $E_{ph}$ and mutual information are:
\begin{equation}
    E_{p}{(A:B)} = \frac{\lads}{4G_{N}} \log{\left( \frac{1+\sin{\alpha}}{1-\sin{\alpha}} \right)}
\end{equation}
\begin{equation}
    \frac{1}{2} I{(A:B)} = \frac{\lads}{2G_{N}} \log{(\tan\alpha)}
\end{equation}
and one can check that the first one is larger than the second. Finally, consider the BTZ black hole, with $A$, $B$ taken to be each half the boundary (on one side). In this case the $E_{ph}$ and the mutual information are given by:
\begin{equation}
    E_{p}{(A:B)} = \frac{\lads}{2G_{N}} \log{\left( \frac{2r_{c}}{r_{+}} \right)}
\end{equation}
\begin{equation}\label{HEPBTZSym}
    \frac{1}{2}I{(A:B)} = \frac{\lads}{2G_{N}} \log{\left(\frac{2r_{c}}{r_{+}}\right)} + \frac{\lads}{2G_{N}} \log{\left( \sinh{\frac{\pi r_{+}}{2\lads}} \right)} - \frac{\pi r_{+}}{4G_{N}}
\end{equation}
To see that $E_{p}{(A:B)} > \frac{1}{2}I{(A:B)}$, we have to argue:
\begin{equation}
    2\lads \log{\left( \sinh{\frac{\pi r_{+}}{2\lads}} \right)} - \pi r_{+} \leq 0
\end{equation}
This is easy to show:
\begin{equation}
    2\lads \log{\left( \sinh{\frac{\pi r_{+}}{2\lads}} \right)} = 2\lads \log{\left( \frac{e^{\pi r_{+}/2\lads} - e^{-\pi r_{+}/2\lads}}{2} \right)} \leq 2\lads \log{e^{\pi r_{+}/2\lads}} = \pi r_{+}
\end{equation}
where we used the fact that the log is a monotonic function. This verifies the bound (\ref{EPLowerBound}).\\

\begin{figure}[H]
    \centering
     \includegraphics[width=7cm]{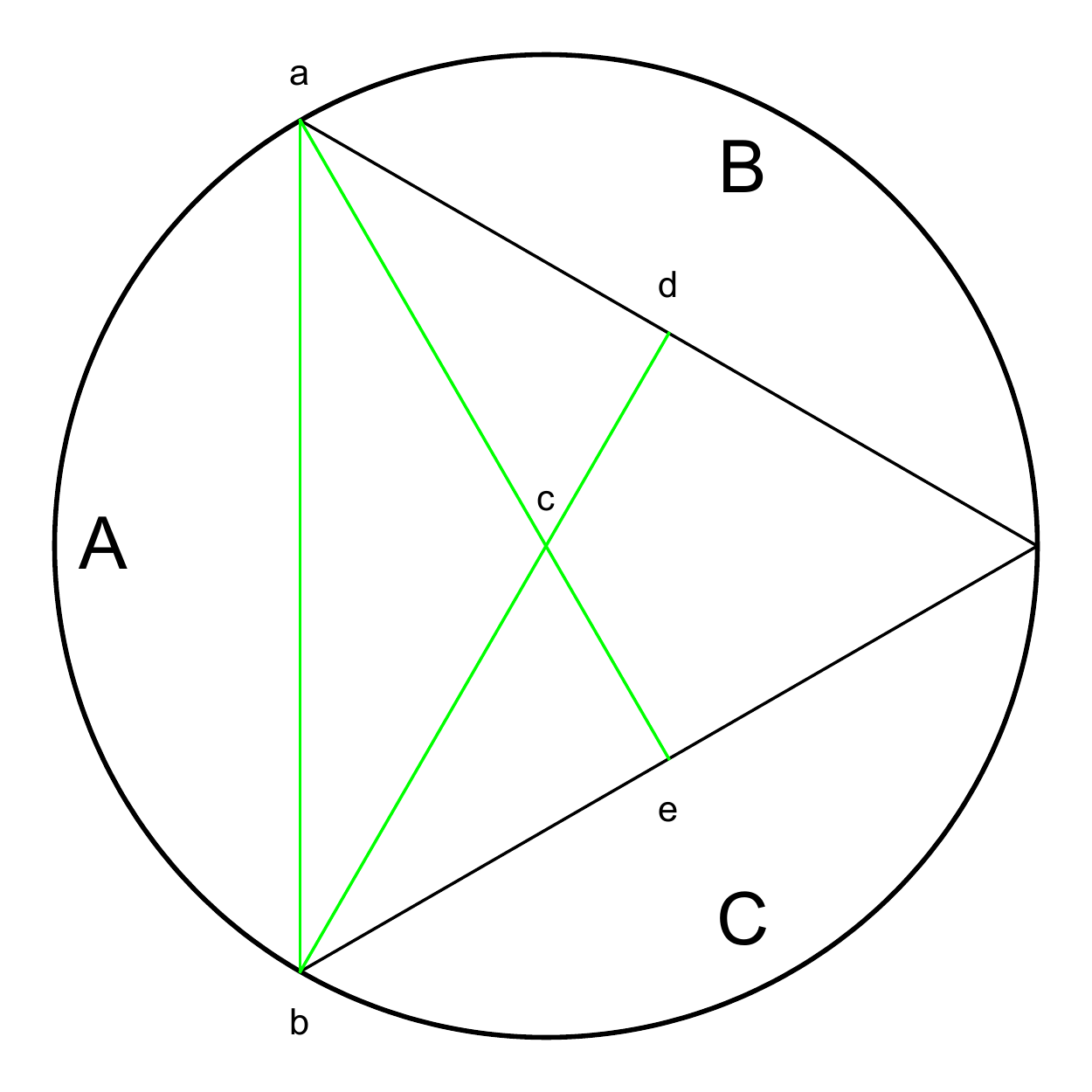}
    \caption{Graphical proof polygamy of $E_{ph}$ for tripartite pure state. Beltrami-Klein coordinates are used here.}
    \label{fig:PolygamyPf}
\end{figure}

\paragraph{Tripartite bound}. Next, consider the tripartite bound (\ref{EPTripartiteBound}). We note a relevant fact: in a holographic state, the mutual information in holographic states is known to be monogamous:
\begin{equation}\label{HolographicMonogamy}
    I{(A:BC)} \geq I{(A:B)} + I{(A:C)}
\end{equation}
as proved in \cite{Hayden:2011ag}. This property combined with the lower bound (\ref{EPLowerBound}) implies the tripartite bound (\ref{EPTripartiteBound}). To see this, let us replace $B$ by $BC$ in the bound (\ref{EPLowerBound}). We obtain:
\begin{equation}
    E_{p}{(A:BC)} \geq \frac{I{(A:BC)}}{2}
\end{equation}
Using the monogamy relation (\ref{HolographicMonogamy}) to replace $I{(A:BC)}$ on the right-hand side then yields the bound (\ref{EPTripartiteBound}). Thus, it will be sufficient to check the bound (\ref{EPLowerBound}) holographically.\\
\paragraph{Polygamy of tripartite pure state.} Finally, we check the polygamy of the $E_{ph}$ for tripartite pure states. Like the lower bound by the mutual information, this property is a simple consequence of Riemannian geometry as illustrated in Figure \ref{fig:PolygamyPf}.
If we denote by $(ab)$ the geodesic length between $a$ and $b$ on this figure etc, then we have:
\begin{equation}
    E_{p}{(A:B)} = \frac{1}{4G_{N}}\left[(ac) + (ce)\right]
\end{equation}
\begin{equation}
    E_{p}{(A:C)} = \frac{1}{4G_{N}} \left[(bc)+(cd)\right]
\end{equation}
\begin{equation}
    E_{p}{(A:BC)} = \frac{1}{4G_{N}} (ab)
\end{equation}
But $(ab) < (ac) + (bc)$ by virtue of being a geodesic. Therefore clearly $E_{p}{(A:B)} + E_{p}{(A:C)} \geq E_{p}{(A:BC)}$. Even though we draw AdS in Figure \ref{fig:PolygamyPf}, it is clear from the proof above that it applies to any asymptotically AdS geometry, and not only empty AdS.\\
\paragraph{Generalization to time-dependent situations}. Finally, we generalize the arguments above for time-dependent backgrounds, starting with the lower bound by half the mutual information. Note that the geometrical argument presented above in the time-independent case does not directly apply due to the fact that in general, the different extremal surfaces involved lie on different spatial slices. However, one can adapt the techniques of \cite{Wall:2012uf} to prove this lower bound, as follows.\\
Consider for instance a spatial slice of the boundary of global AdS, and let $A$ and $B$ be ``large'', non-adjacent boundary intervals (we require them to be large so that the $E_{ph}$ is nonzero). By corollary (h) of Theorem 17 in \cite{Wall:2012uf}, we know that there exists a spatial slice $\Sigma$ containing the HRT surfaces for $A$, $B$ and $AB$, and on which all these HRT surfaces are minimal. Thus, one can draw a picture analogous to the left panel of Figure \ref{fig:PlotEP}, except that the spatial slice shown is $\Sigma$ and not a static time slice. The green curve on this Figure is now taken to be the minimal curve \textit{lying on $\Sigma$} which connects the two components of the HRT surface for $S(AB)$. Note, in particular, that this green curve does not in general compute the $E_{ph}$ since the curve that does is not confined to the slice $\Sigma$. However, by the minimax property of extremal surfaces shown in \cite{Wall:2012uf}, we know that the green curve is shorter in length than the curve computing the $E_{ph}$. This fact, combined with the same argument for the lower bound in the static case but repeated on the slice $\Sigma$, establishes the lower bound in time-dependent settings: $E_{ph}{(A:B)} \geq \frac{1}{2}I{(A:B)}$.\\
The tripartite bound $I{(A:BC)} \geq I{(A:B)} + I{(A:C)}$ also holds in the time-dependent case since the monogamy of mutual information is known to be true (with the assumption of null curvature condition). This is, again, established in \cite{Wall:2012uf}.

\section{Conclusion and future work}

We presented an analysis of the entanglement of purification in three different model many-body systems. In the case of random stabilizer tensor networks we were able to actually compute the entanglement of purification. Our holographic calculations focused for simplicity on the case of a three dimensional bulk, but the proposal obviously extends to any dimension. One technical challenge is to show that the desired properties of $E_p$ are obeyed by our holographic proposal in the time dependent case. We found reasonably good agreement between the holographic results and a numerical study of the Ising spin chain.

We mention two promising directions for future work within holography: (1) exploring the connection between the $E_{p}$ and the differential entropy \cite{Czech:2014tva} as well as kinematic space, and (2) exploring the connection between $E_{p}$ and the bit threads \cite{Freedman:2016zud}. It has been discovered that the lengths of arbitrary curve in the bulk can be interpreted by terms of quantum information by a quantity called the differential entropy. This latter quantity is associated to a continuum of boundary intervals defined by the family of geodesics in the bulk tangential to the curve of interest. Equivalently, the length of curves can also be computed by integrating over the volume of a region in an auxiliary geometry called kinematic space. Remarkably, volume elements in kinematic space turn out to compute the conditional mutual information of 3 adjacent boundary intervals. Of course, the differential entropy/kinematic space interpretation also applies to the geodesic segments computing the $E_{ph}$. Therefore, there seems to be deep connection between holographic entanglement of purification and other quantum-information-theoretical quantities such as the conditional mutual information.\\

On the other hand, the Ryu-Takayanagi formula has been reinterpreted recently via the min-cut/max-flow theorem as some kind of information flow \cite{Freedman:2016zud}. Within this framework, a beautiful picture emerges for the lower bound of the $E_{ph}$ by half the mutual information, as follows: one can construct a flow in the bulk that computes half the mutual information and which is supported only in the entanglement wedge. The $E_{ph}$ then acts as the bottleneck that restricts this flow, in pretty much the same way as the diameter of a pipe contrains the amount of water flowing across it. Further explorations of this bit thread picture may help prove nontrivial properties of the $E_{ph}$ that are not easily seen otherwise.

In the context of spin chains, we have shown that a substantial reduction in entanglement relative to the thermofield double state is possible. One promising direction is to construct new tensor network algorithms that take some advantage of this potential reduction in entanglement. Finding the right balance between the cost of keeping unneeded entanglement and the cost of finding and removing it is an interesting challenge.

Finally, in the context of tensor network model of holography, we computed the entanglement of purification for random stabilizer tensor networks. Despite the fact that these networks obey the discrete RT formula, the discrete analog of the holographic proposal for $E_p$ was actually not obeyed in general. This is presumably due to the rather simple structure of entanglement in these networks. It would be very interesting to study random tensor networks, for example, to see if the analog of $E_{ph}$ does actually compute $E_p$ in that case.

\textit{Acknowledgements:} MPZ is particularly indebted to Frank Pollmann, with whom we first considered and implemented the purification disentangling algorithm. BGS acknowledges support from the Simons Foundation as part of the It From Qubit Collaboration. This work is partially supported by NSF grants PHY1407744 and PHY1708139. MPZ  acknowledges the hospitality of the Aspen Center for Physics, which is supported by National Science Foundation grant PHY-1607611.

\appendix

\section{Proofs of properties of $E_{p}$}\label{App:Proofs}
In this appendix, we review the proofs of the properties of the $E_{p}$ mentioned in the Introduction \cite{2002JMP....43.4286T,2015PhRvA..91d2323B}, starting with the upper bound (\ref{EPUpperBound}) by the entanglement entropy.\\
\textit{Proof}: Let $\rho_{AB}$ be a bipartite density matrix with eigenvalues $\lambda_{i}$ and eigenvectors $| \psi_{i} \rangle$. The standard purification of $\rho_{AB}$:
\begin{equation}
    |\psi \rangle = \sum_{i} \sqrt{\lambda_{i}} |\psi_{i}\rangle_{AB} \times |0\rangle_{A'} |i\rangle_{B'}
\end{equation}
yields the entanglement entropy $S{(A)}$ when we trace out the $BB'$, and $S(B)$ when we trace out the $AA'$. Since we have to minimize over all purifications in the definition of the EP, the bound (\ref{EPUpperBound}) follows.\\
Next, we prove monotonicity (\ref{Monotonicity}).\\
\textit{Proof}: Let $\rho_{ABC}$ be the density matrix on $ABC$. If $\rho_{ABC}$ is pure, then the EP coincides with the entanglement entropy: $E_{p}{(A:BC)} = S(A)$. But $E_{p}{(A:B)}$ is bounded above by $S(A)$, hence monotonicity is satisfied. If $\rho_{ABC}$ is mixed, then we note that the set of purifications of the form $|\psi \rangle_{AA';(BC)(BC)'}$ is a subset of the purifications of $\rho_{AB}$ of the form $|\psi \rangle_{AA';BB'}$, and monotonicity follows immediately.\\
Next, we prove the lower bound (\ref{EPLowerBound}) by the mutual information.\\
\textit{Proof}: Let $| \psi \rangle_{ABA'B'}$ be the optimal pure state for the evaluation of $E_{p}{(A:B)}$, i.e. $S(AA',|\psi\rangle) = E_{p}{(A:B)}$. USing the subadditivity of the conditional entropy for a 4-party quantum state:
\begin{equation}
    S{(A'B'|AB)} \leq S{(A'|A)} + S{(B'|B)}
\end{equation}
Using the definition of conditional entropy ($S(A|B)=S(AB)-S(B)$), this implies:
\begin{equation}
    S{(ABA'B')} - S{(AB)} \leq S{(AA')} - S{(A)} + S{(BB')} - S{(B)}
\end{equation}
But $S{(ABA'B')} = 0$ since $\rho_{ABA'B'}$ is pure by definition of the EP, and $S{(AA')} = S{(BB')} = E_{p}{(A:B)}$. The above simplifies to:
\begin{equation}
    S{(A)} + S{(B)} - S{(AB)} \leq 2E_{p}{(A:B)}
\end{equation}
which is equivalent to (\ref{EPLowerBound}).\\
Next, we prove the lower bound (\ref{EPTripartiteBound}) for the tripartite systems.\\
\textit{Proof}: Let $|\psi \rangle_{ABCA'D'}$ be the optimal pure state for evaluating the EP, i.e.
\begin{equation}
    E_{p}{(A:BC)} = \frac{1}{2}I{(AA':BCD')}
\end{equation}
We now use the fact that mutual information satisfies the monogamy equality condition for pure states:
\begin{equation}
    I{(AA':BCD')} = I{(AA':B)} + I{(AA':CD')}
\end{equation}
to obtain
\begin{equation}
    E_{p}{(A:BC)} = \frac{1}{2}I{(AA':B)} + \frac{1}{2}I{(AA':CD')}
\end{equation}
But the mutual information is monotonic, i.e. $I{(AA':B)} \geq I{(A:B)}$ and $I{AA':CD'} \geq I{(A:C)}$. The bound (\ref{EPTripartiteBound}) then follows.\\
Next, we show that the $E_{p}$ in a state saturating the Araki-Lieb inequality is the entanglement entropy of the smaller subsystem.\\
\textit{Proof}: Saturation of Araki-Lieb means:
\begin{equation}
    S{(A)} - S{(B)} = S{(AB)}
\end{equation}
Note that the EP is bounded above by the entanglement entropy and below by half the mutual information:
\begin{equation}
    \frac{1}{2}I{(A:B)} \leq E_{p}{(A:B)} \leq S{(B)}
\end{equation}
But Araki-Lieb saturation also implies $I{(A:B)} = 2S{(B)}$. The above becomes:
\begin{equation}
    S{(B)} \leq E_{p}{(A:B)} \leq S{(B)}
\end{equation}
Hence $E_{p}{(A:B)}=S{(B)}$.\\
Next, we show that the $E_{p}$ in a tripartite pure state is polygamous (inequality \ref{EPPolygamy}).\\
\textit{Proof}: By the lower bound by the mutual information $E_{p}{(A:B)} \geq \frac{I{(A:B)}}{2}$, we have:
\begin{equation}
    E_{p}{(A:B)} + E_{p}{(A:C)} \geq \frac{1}{2} (I(A:B) + I(A:C))
\end{equation}
Recall that in a pure state, the mutual information satisfies the monogamy equality $I(A:B)+I(A:C)=I(A:BC)=S(A)$. But $S(A)= E_{p}{(A:BC)}$ since the state is pure. Thus,
\begin{equation}
    E_{p}{(A:B)} + E_{p}{(A:C)} \geq E_{p}{(A:BC)}
\end{equation}
Finally, we show that for a classically correlated state of the form $\rho_{AB} = \sum_{i} p_{i} |i\rangle \langle i|_{A} \otimes |i\rangle \langle i|_{B}$, the $E_{p}$ is the Shannon entropy of the corresponding probability distribution: $E_p = -\sum_i p_i \log p_i$.\\
\textit{Proof}: We copy the classical information to a third system $C$ and consider the state:
\begin{equation}
    \rho_{ABC} = \sum_{i} p_{i} |i\rangle \langle i|_{A} \otimes |i \rangle \langle i|_{B} \otimes |i\rangle \langle i|_{C}
\end{equation}
This state is unitarily related to the state $\rho_{AB}$. Indeed, if we call $V$ a unitary operator that copies the classical information $V |i\rangle_{B} |0\rangle_{C} = |i\rangle_{B} |i\rangle_{C}$ for some reference state $| 0 \rangle_{B}$, we then have:
\begin{equation}
    \rho_{ABC} = V \rho_{AB} \otimes |0\rangle \langle 0|_{C} V^{\dagger}
\end{equation}
Using the inequalities previously established in this appendix, we have:
\begin{equation}
    S{(A)} \geq E_{p}{(A:B)} = E_{p}{(A:BC)} \geq \frac{1}{2}I{(A:B)} + \frac{1}{2} I{(A:C)}
\end{equation}
But $S{(A)} = I{(A:B)} = I{(A:C)} = -\sum_{i} p_{i} \log{p_{i}}$. Thus, we have:
\begin{equation}
    E_{p}{(A:B)} = -\sum_{i} p_{i}\log{p_{i}}
\end{equation}

\section{Shortest distance between 2 geodesics via Beltrami-Klein coordinates}\label{App:Distance}
In this appendix, we use the Beltrami-Klein model of the hyperbolic plane \cite{Beltrami1,Beltrami2} together with its well-known properties to compute the shortest distance between any two geodesics in the hyperbolic plane $\mathbb{H}^{2}$. To this effect, we use the following fact (also known as the \textit{ultraparallel theorem} in hyperbolic geometry):\\

\textbf{Fact}. \textit{Given any two geodesics in the hyperbolic plane which do not share a common endpoint on the boundary (i.e. given two ultra-parallel curves), then there exists a unique geodesic which is perpendicular to both of them. Moreover, this common perpendicular is the shortest curve between the two given geodesics.}\\

\begin{figure}[H]
    \centering
    \includegraphics[width=5cm]{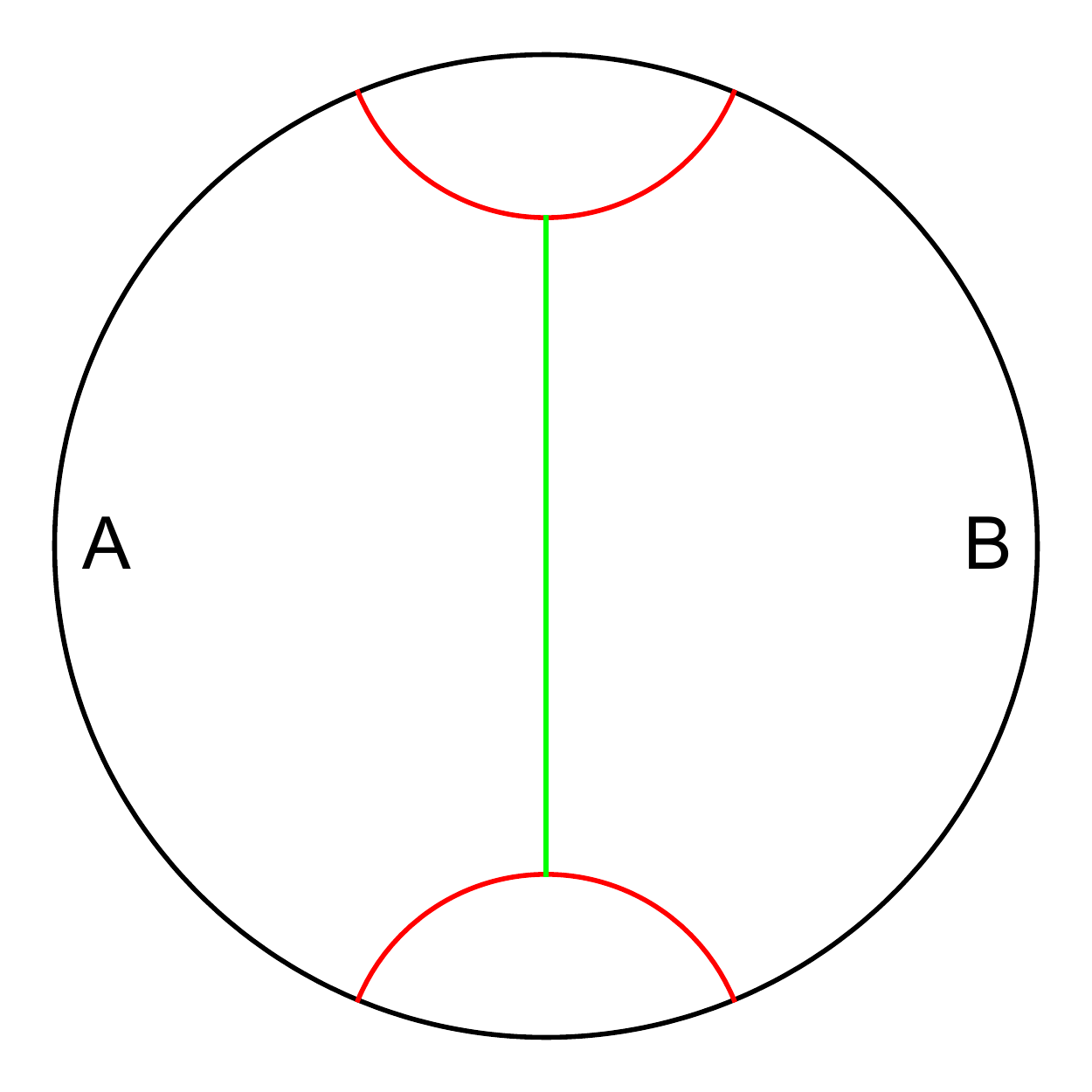}
    \includegraphics[width=5cm]{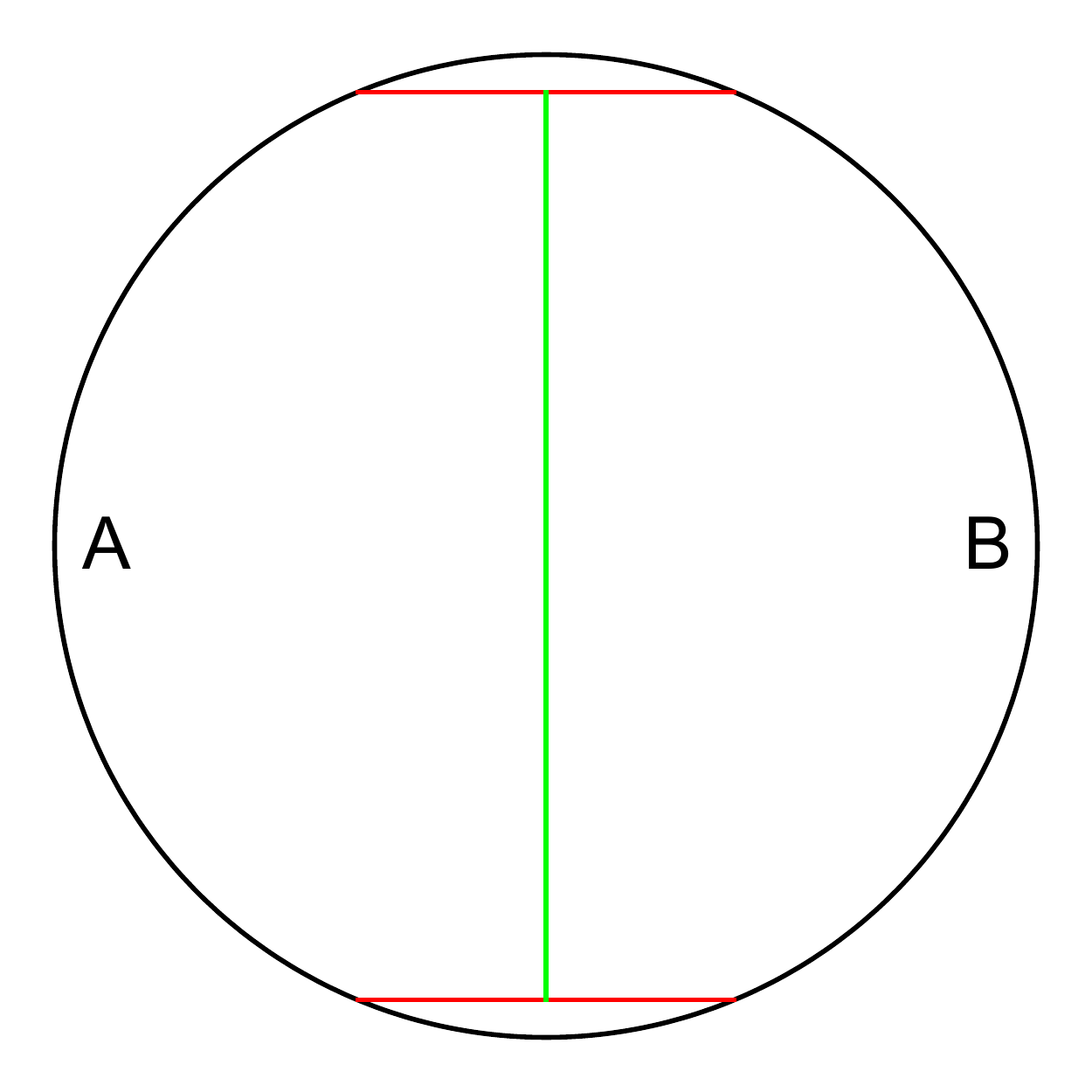}
    \caption{Left: Plot of the RT surface (red) and the EP surface (green) in the Poincar\'{e} disk model. Right: the same plot as it appears in the BK model}
    \label{fig:EPBK}
\end{figure}

By the fact above, we should construct the unique common perpendicular to the two given geodesics in order to find the shortest distance between them. We will work with the Beltrami-Klein (BK) model of the hyperbolic plane to construct the common perpendicular. The BK metric can be obtained from the usual global coordinates in AdS by a redefinition of the radial coordinate:
\begin{equation}
    \frac{r}{\lads} = \frac{R}{\sqrt{R^{2}+\lads^{2}}}
\end{equation}
In the BK model, geodesics are straight lines. For example, in Figure \ref{fig:EPBK} we draw the RT surface as well as the EP surface for the case where $A$ and $B$ are of the same size and diametrically opposite from each other, both in the Poincar\'{e} disk model and BK model.
In the simple case of Figure \ref{fig:EPBK}, the unique common perpendicular is easily seen to be the line connecting the midpoints of the two red lines (by symmetry). For more general boundary intervals $A$ and $B$, finding the common perpendicular is a bit more involved, but the following fact is helpful:\\

\textbf{Fact}. \textit{Let $L$ be a geodesic in the hyperbolic plane. Another geodesic $L'$ is perpendicular to $L$ if and only if it goes throught the pole of $L$ when extended beyond the edge of the disk (in the Beltrami-Klein model). Here the pole of $L$ is the intersection between the two lines tangential to the edge of the disk at the two endpoints of $L$.}\\

Using the fact above, we can then construct the common perpendicular to any two geodesics as in Figure \ref{fig:PlotCommonPerp} below. Let $a$, $b$, $c$, $d$ be 4 boundary points, and we have two geodesics $\mathcal{L}_{1}$ and $\mathcal{L}_{2}$ connecting $a$ to $b$ and $c$ to $d$ respectively. These two geodesics are black lines in Figure \ref{fig:PlotCommonPerp}. By the fact above, we know that the unique common perpendicular to $\mathcal{L}_{1}$ and $\mathcal{L}_{2}$ passes through the poles of both $\mathcal{L}_{1}$ and $\mathcal{L}_{2}$. The pole of $\mathcal{L}_{1}$ is the point $p$, which is the intersection of the two tangential lines to the disk at $a$ and $b$ (depicted in red, dashed in the Figure). Similarly, the pole of $\mathcal{L}_{2}$ is the point $q$. The green line connecting $p$ to $q$ is then the unique commone perpendicular to $\mathcal{L}_{1}$ and $\mathcal{L}_{2}$.\\
Let $m$ and $n$ be the intersection of the green line with $\mathcal{L}_{1}$ and with $\mathcal{L}_{2}$ respectively, and let $r$ and $s$ be the two intersections of the green line with the edge of the disk. The shortest distance between $\mathcal{L}_{1}$ and $\mathcal{L}_{2}$ is then the distance between $m$ and $n$. Using the standard formula for distance in the Beltrami-Klein model:
\begin{equation}\label{DistanceBK}
    d{(m,n)} = \frac{\lads}{2} \log{\frac{|sm||nr|}{|sn||mr|}}
\end{equation}
where $|\cdot|$ is the Euclidean distance between the two points. Note that the distance is a function of a the cross-ratio of the 4 points. Our task now is to relate the 4 points $m$, $n$, $r$ and $s$ in the formula above to the 4 points $a$, $b$, $c$, $d$. Let us denote by $\alpha_{1}$, $\alpha_{2}$ the half-widths of $(a,d)$ and $(b,c)$ respectively, and by $\phi_{1}$, $\phi_{2}$ the midpoints of $(a,d)$ and $(b,c)$. Note that the intervals we are referring to are \textit{not} $(ab)$ and $(cd)$ but the other two. We want to write down a formula for $d{(\phi_{1},\alpha_{1},\phi_{2},\alpha_{2})}$. After some analytical geometry, we find the formula (\ref{EphNonAdjacent}) for $E_{ph}$ of 2 non-adjacent intervals.\\
\begin{figure}[H]
    \centering
    \includegraphics[width=8cm]{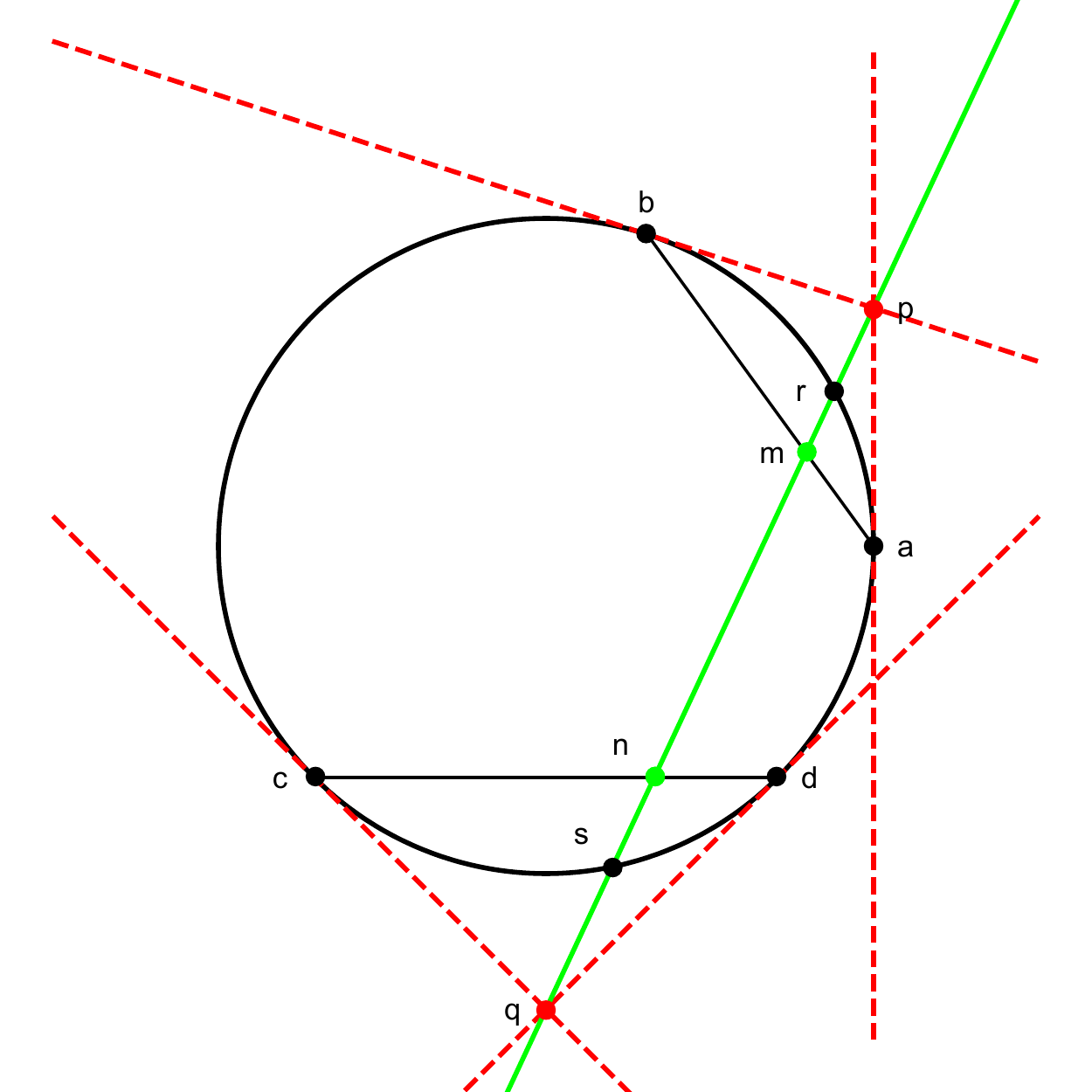}
    \caption{The RT surface is in red. The EP surface is in green.}
    \label{fig:PlotCommonPerp}
\end{figure}
Next, we consider the limiting case where one of the two geodesics shrinks to a point on the boundary. Of course, the distance between the remaining geodesic and the point on the boundary is divergent and we have to regularize it. The shortest curve from the geodesic to the point can be constructed using the techniques previously described: by constructing the line going through the pole of the geodesic to the point on the boundary (see Figure \ref{fig:PlotAdjacent} below).
\begin{figure}[H]
    \centering
    \includegraphics[width=7cm]{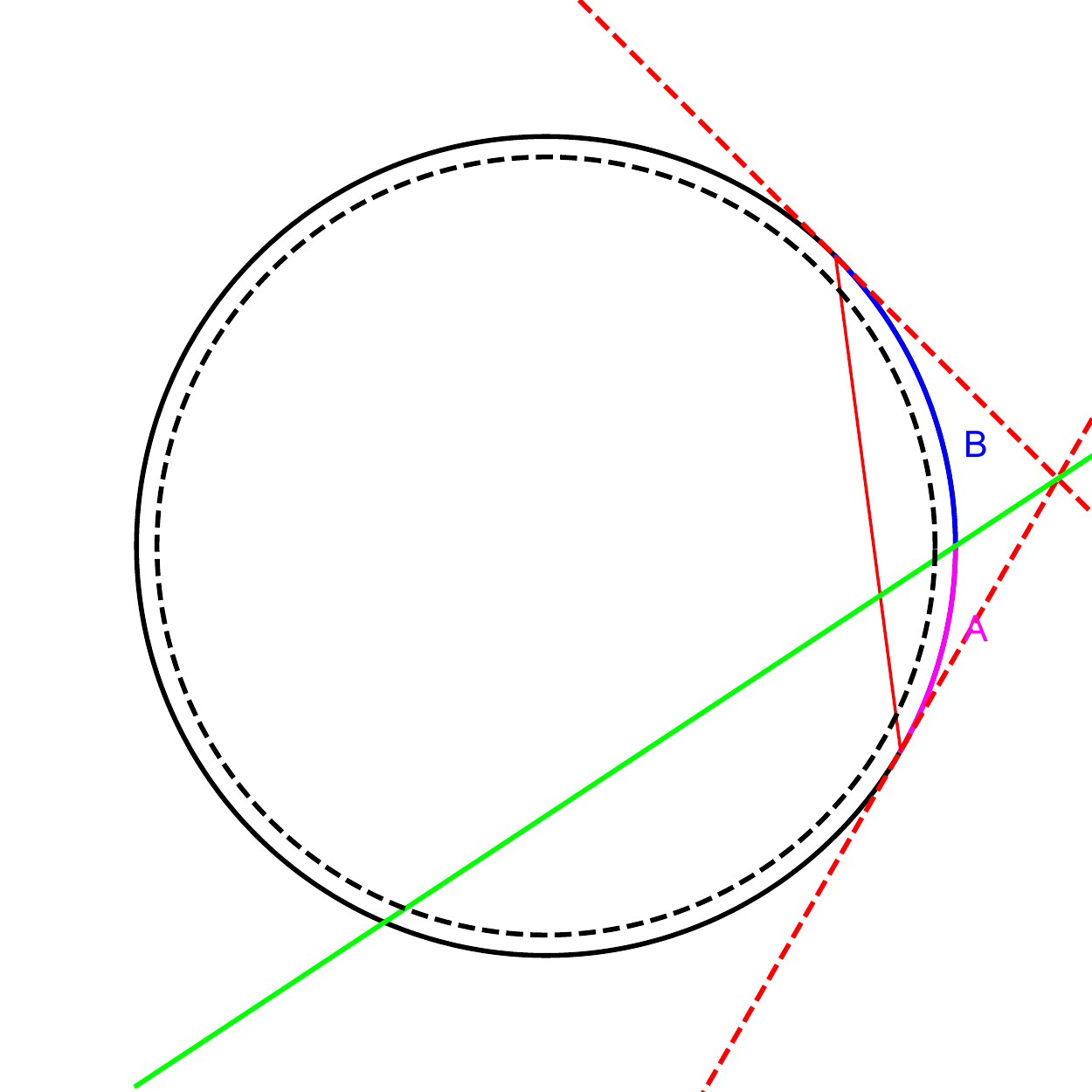}
    \caption{The RT surface for $AB$ is in red. The EP surface for $E_{p}{(A:B)}$ is in green. The regularizing surface is in black dashed.}
    \label{fig:PlotAdjacent}
\end{figure}
Unlike the non-adjacent case, the EP is now divergent. We regularize it length by introducing a cutoff at radius $\lads(1-\epsilon)$ (dashed circle in the figure above). Thus, we want to compute the length of the green line segment between the dashed circle and the RT surface. As in the non-adjacent case, we parametrize $A$ and $B$ as $(\phi_{1}-\alpha_{1},\phi_{1}+\alpha_{1})$ and $(\phi_{2}-\alpha_{2},\phi_{2}+\alpha_{2})$ respectively. The fact that they are adjacent implies:
\begin{equation}
    \phi_{2} = \phi_{1} + \alpha_{1} + \alpha_{2}
\end{equation}
After some analytical geometry, we find the distance formula:
\begin{equation}
    d = \frac{1}{2} \log{\left[ \frac{g^{2}}{\tan^{2}{(\alpha_{1}+\alpha_{2})}} \frac{(2\sin{\alpha_{1}}\sin{\alpha_{2}} + \cos{(\alpha_{1}+\alpha_{2})}\sqrt{g^{2}(1-\epsilon)^{2}-\sec^{2}{(\alpha_{1}+\alpha_{2})}\sin^{2}{(\alpha_{1}-\alpha_{2})}}} {(2\sin{\alpha_{1}}\sin{\alpha_{2}} - \cos{(\alpha_{1}+\alpha_{2})}\sqrt{g^{2}(1-\epsilon)^{2}-\sec^{2}{(\alpha_{1}+\alpha_{2})}\sin^{2}{(\alpha_{1}-\alpha_{2})}}} \right]}
\end{equation}
If we now expand in $\epsilon$ around $\epsilon=0$, we find the result (\ref{EphAdjacent}) given in section \ref{Sec:Holography}.

\section{Minimization of 2nd Renyi Entropy}\label{app:s2min}

    In this appendix we describe the disentangling step of the numerical calculation described in Section~\ref{sec:numerics} in more detail.

    The disentangling step seeks to efficiently find a unitary transformation on the ancilla degrees of freedom of our system which minimizes the total entropy.
    While this unitary could be any global unitary transformation, to make the problem tractable we instead sweep across the system, minimizing the Second Renyi Entropy between two sites at a time.
    Disentangling algorithms are discussed in more detail in Ref.~\cite{vidalmera}.

\begin{figure}
        \centering
        \includegraphics[width=0.5\textwidth]{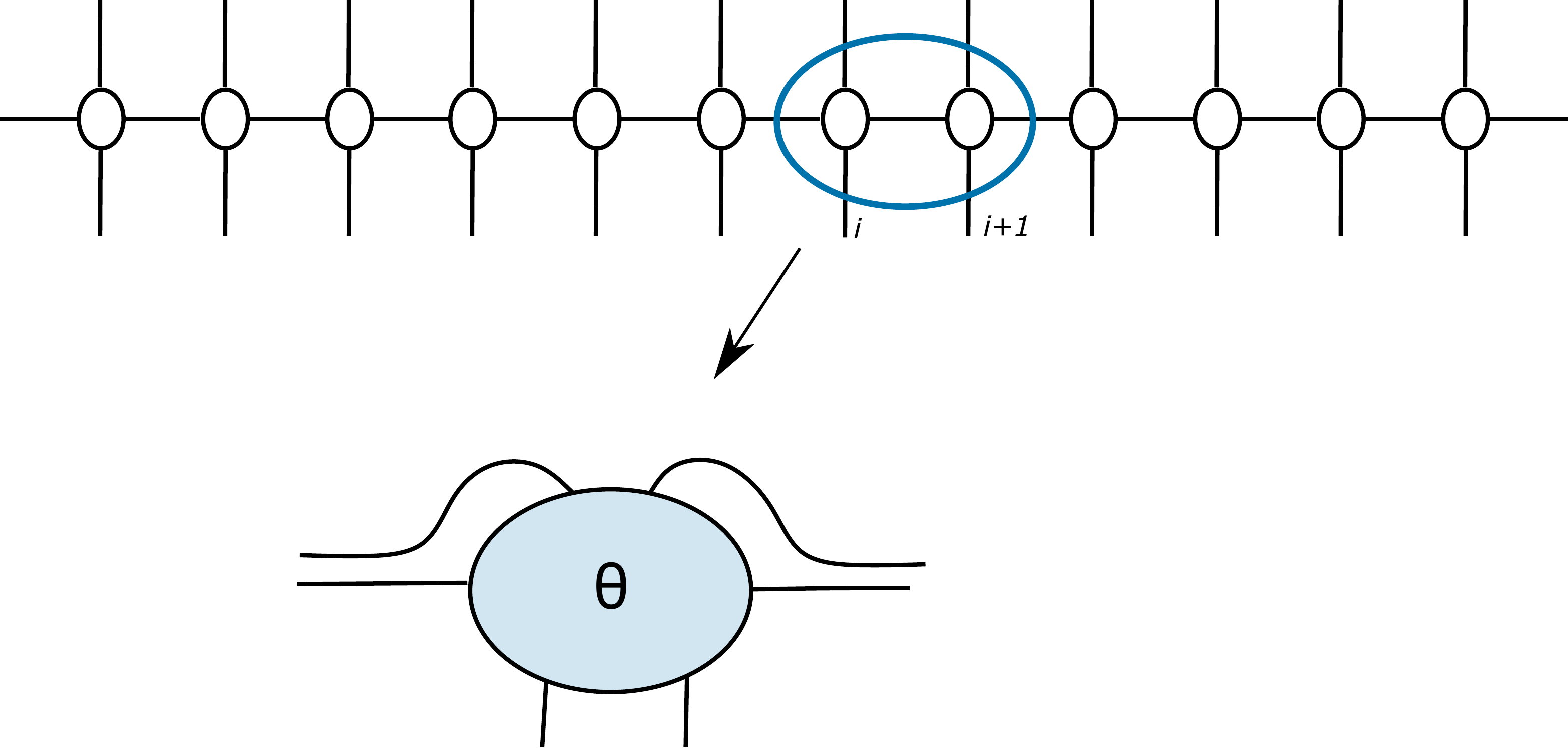}
        \caption{ The state~$\Theta$. We combine the bond and physical degrees of freedom into a pair of physical indices represented by the horizontal legs.
        The ancilla degrees of freedom (the bottom legs) are acted upon by our two-site disentangler.
    }%
    \label{fig:theta}
   \end{figure}

    Once the center of normalization for the MPS is on site~$i$ or~$i+1$, the state can be represented by the object~$\Theta$\cite{schollwock}, depicted in Figure~\ref{fig:theta}.
    We calculate the Second Renyi Entropy~$S_2 = -\log\text{Tr}\rho^2$ in the usual way, treating~$\Theta$ as our state.

\begin{figure}
        \centering
        \includegraphics[width=0.5\textwidth]{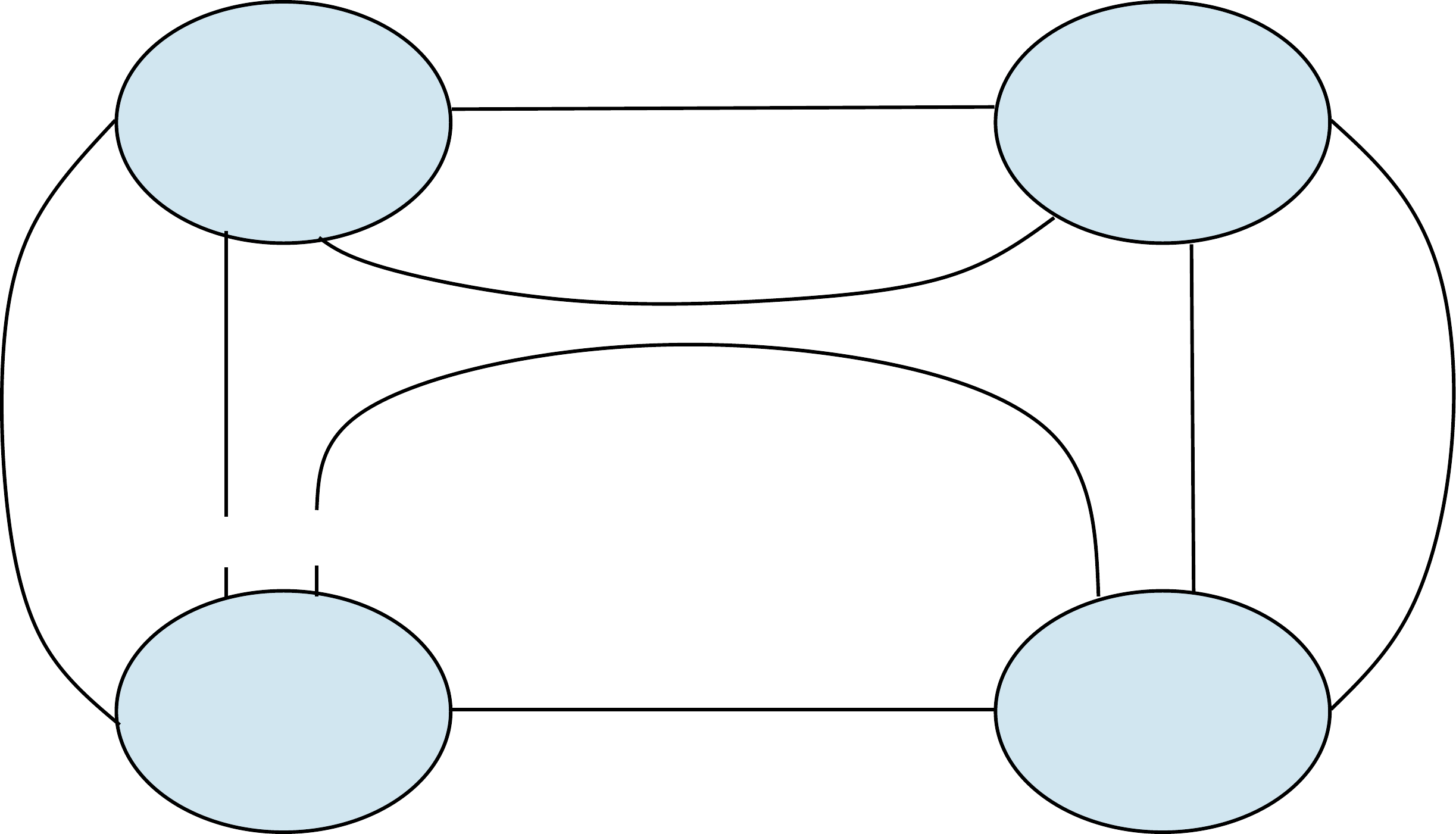}
        \caption{An illustration of the gradient operator~$\frac{\partial \text{Tr} \rho^2}{\partial U}$ evaluated at the identity.
        Each oval represents~$\Theta$ or its conjugate.
        Every pair of connected ancilla legs is connected by the identity, while the disconnected set of legs represents the removed unitary transformation.
    }%
    \label{fig:gradient}
    \end{figure}

     To minimize this quantity for our pair of sites, we use a modified steepest descent algorithm.
    In particular, we apply a unitary disentangler to the ancilla legs of~$\Theta$, and express~$S_2$ in terms of this unitary.
    We then calculate the gradient of~$\text{Tr}\rho^2$ with respect to this unitary, evaluated at the identity.
    This gradient is depicted graphically in Figure~\ref{fig:gradient}.

    The algorithm then chooses a unitary disentangler close to this gradient, which we obtain via a singular value decomposition.
    In particular, for the decomposition
    \begin{equation}\label{eq:gradSVD}
        \frac{\partial\text{Tr}\rho^2}{\partial U} = XYZ\, ,
    \end{equation}
    where~$X$ and~$Z$ are unitary matrices, the two-site disentangler chosen by the algorithm is~$U^\prime = XZ$. This selects the unitary closest to~$XYZ$, as defined by the matrix norm.

\begin{figure}
    	\centering
    	\includegraphics[width=0.5\textwidth]{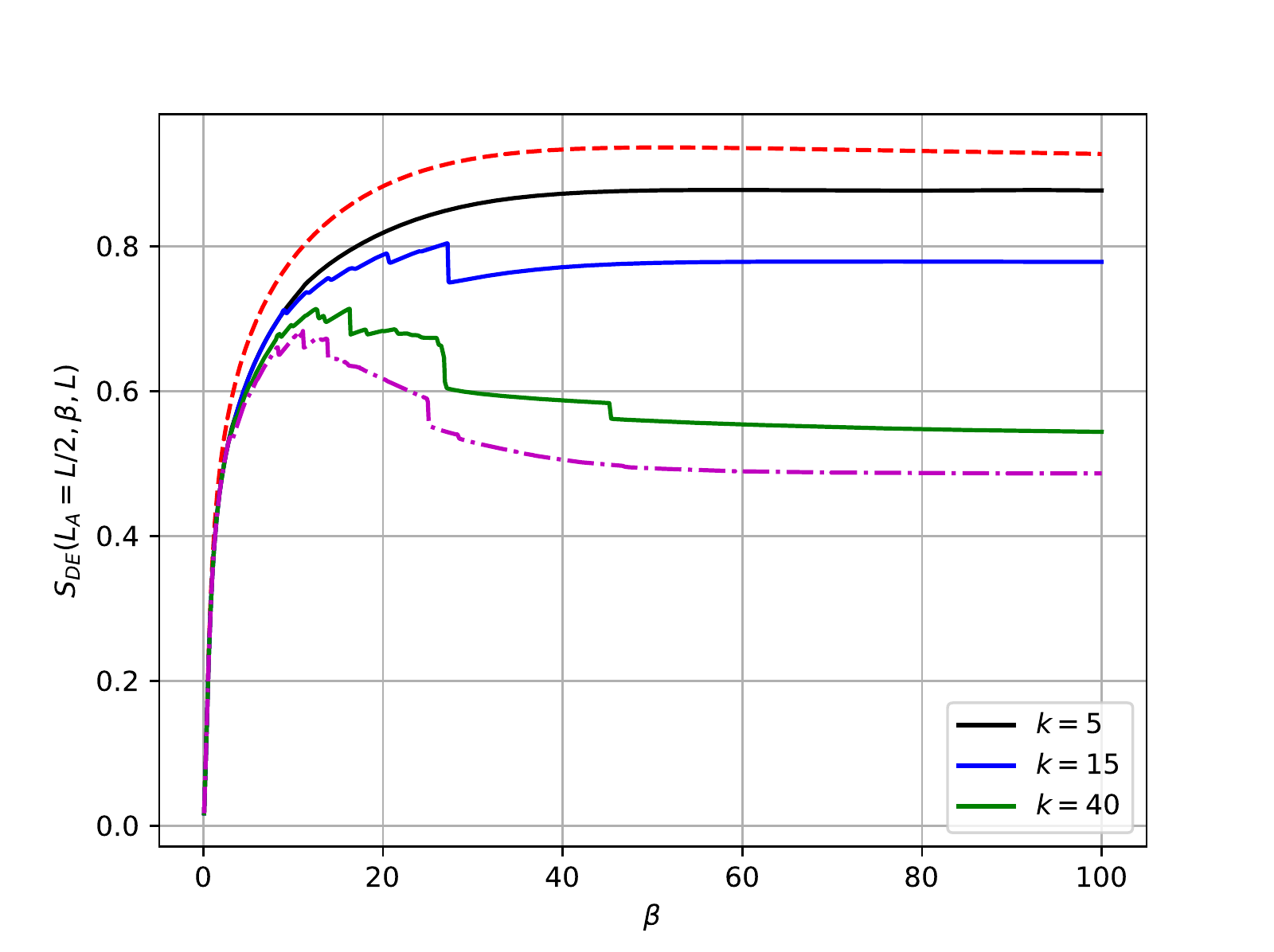}
    	\caption{The disentangled entropy~$S_{\text{DE}}$ at the middle cut for different choices of two-site disentangler given by~\eqref{eq:altSVD}, for a system size~$L=30$.
    	The dashed line shows the entropy of the TFD state without disentangling, while the dash-dotted line is the disentangled entropy using the prescription~\eqref{eq:gradSVD}.
    	}%
    \label{fig:noise}
    \end{figure}

    As argued in Section~\ref{sec:numerics}, this approach does well to approximate the entanglement of purification, but the data contains considerable noise for intermediate values of~$\beta$.
    One method to reduce the noise is to choose two-site disentanglers which are closer to the identity.
    For example, an alternate approach would be to instead choose the decomposition
    \begin{equation}\label{eq:altSVD}
        1 + k\frac{\partial \text{Tr}\rho^2}{\partial U} = XYZ\, ,
    \end{equation}
    for a small value of~$k$, with~$U^\prime = XZ$ as before.
    This choice of disentangler corresponds to the standard steepest descent algorithm (again with the restriction that only unitary disentanglers are allowed).
    The choice~\eqref{eq:gradSVD} corresponds to the large~$k$ limit of~\eqref{eq:altSVD}.
    Figure~\ref{fig:noise} shows the entropy after disentangling using various values of~$k$.

    Unforunately, small values of~$k$ lead to sub-optimal disentanglers, as the algorithm converges on local minima more readily when~$k$ is small.
    As Figure~\ref{fig:noise} suggests, the noise becomes significant once the algorithm is able to escape some local minima, even for suboptimal purifications.
    This suggests that the noise is in part due to movement between local minima.
    Escaping these local minima, however, appears essential to produce a good approximation of the entanglement of purification, as we argue our algorithm accomplishes.

\end{document}